\documentclass[11pt]{article}
\usepackage{cite}
\usepackage{amsmath,amsfonts,amssymb}
\usepackage[small,bf,hang]{caption}

\usepackage{slashed}
\usepackage[normalem]{ulem}

\def\hybrid{
        \topmargin -20pt
        \oddsidemargin 0pt
        \headheight 0pt \headsep 0pt
        \textwidth 6.25in 
        \textheight 9.5in 
        \marginparwidth .875in
        \parskip 5pt plus 1pt \jot = 1.5ex}

\hybrid

 \csname
@addtoreset\endcsname{equation}{section}

\newcommand{\nin}[1] {\underline{\phantom{h}}\hskip-6pt {#1}}

\def\moth{\mathsurround=0pt}
\newdimen\zo \zo=0pt

\def\tick{\leaders\hrule height 0.5ex depth 0pt \hskip 0.5pt}
\def\upboxfill{$\moth \setbox\zo\hbox{\tick}%
  \hskip 3pt\hbox to 0pt{$\tick$\hss}\hrulefill \hbox to 7.5pt{$\tick$\hss}$}

\def\dtick{\leaders\hrule height .34pt depth 0.5ex \hskip 0.5pt}
\def\downboxfill{$\moth \setbox\zo\hbox{\dtick}%
  \hskip 2pt\hbox to 0pt{$\dtick$\hss}\hrulefill \hbox to 2pt{$\dtick$\hss}$}

\def\bec{\begin{center}}
\def\ec{\end{center}}

\def\be{\begin{equation}}
\def\ee{\end{equation}}
\def\bea{\begin{eqnarray}}
\def\eea{\end{eqnarray}}
\def\ba{\begin{array}}
\def\ea{\end{array}}

\begin{document}

\begin{titlepage}
\rightline{}
\rightline{\tt MIT-CTP-4331}
\rightline{\tt  LMU-ASC 75/11}
\rightline{December 2011}
\begin{center}
\vskip 2.5cm
{\Large \bf {
On the Riemann Tensor in Double Field Theory
}}\\
\vskip 2.5cm
{\large {Olaf Hohm${}^1$ and Barton Zwiebach${}^2$}}
\vskip 1cm
{\it {${}^1$Arnold Sommerfeld Center for Theoretical Physics}}\\
{\it {Theresienstrasse 37}}\\
{\it {D-80333 Munich, Germany}}\\
olaf.hohm@physik.uni-muenchen.de
\vskip 0.7cm
{\it {${}^2$Center for Theoretical Physics}}\\
{\it {Massachusetts Institute of Technology}}\\
{\it {Cambridge, MA 02139, USA}}\\
zwiebach@mit.edu

\vskip 1.5cm
{\bf Abstract}
\end{center}

\vskip 0.4cm

\noindent
\begin{narrower}

Double field theory provides T-duality covariant generalized
tensors that are natural extensions of the scalar and Ricci
curvatures of Riemannian geometry.  We search for
a similar extension of the Riemann curvature tensor by
developing a geometry based on the generalized metric and
the dilaton.  We find a duality covariant Riemann tensor whose
contractions give the Ricci and scalar curvatures, but that  is not fully determined in terms of the  physical fields.  This suggests that
$\alpha'$ corrections to the effective action require
$\alpha'$ corrections
to  T-duality transformations and/or  generalized
diffeomorphisms.  Further evidence to
this effect is found by an additional computation that shows that
there is no T-duality invariant four-derivative object built
from the generalized metric and the dilaton that reduces to the
square of the Riemann tensor.

\end{narrower}

\end{titlepage}

\newpage

\tableofcontents

\section{Introduction}

Among the celebrated dualities of string theory T-duality is arguably the
most intriguing, for it directly hints at novel geometrical structures,
transcending the usual framework of differential geometry.
Recently, a so-called double field theory has been constructed that manifests
some of these features at the level of space-time theories for the
massless sector of string theory. Specifically, here the space-time coordinates
are doubled in order to realize the `T-duality group' $O(D,D)$ geometrically,
while introducing an $O(D,D)$ covariant constraint that locally removes the
dependence on half of the coordinates \cite{Hull:2009mi,Hohm:2010jy,Hohm:2010pp}.
(See \cite{Siegel:1993th,Tseytlin:1990nb,Duff:1989tf,Hohm:2010xe,Kwak:2010ew,Hohm:2011gs,Hohm:2011dz,Hohm:2011ex,
Hohm:2011zr,arXiv1108.4937,arXiv1111.7293,Hillmann:2009ci,Berman:2010is,West:2010ev,Jeon:2010rw,Jeon:2011cn,Jeon:2011vx,Schulz:2011ye,
Copland:2011yh,Thompson:2011uw,Albertsson:2011ux,Andriot:2011uh,Coimbra:2011nw}
for previous work and further developments.)

The formulation of double field theory that is perhaps the most intuitive
and which will be used throughout this paper is the generalized metric
formulation. The generalized metric ${\cal H}_{MN}$
 is the $O(D,D)$-valued symmetric tensor
 \be\label{firstH}
  {\cal H}_{MN} \ = \  \begin{pmatrix}    g^{ij} & -g^{ik}b_{kj}\\[0.5ex]
  b_{ik}g^{kj} & g_{ij}-b_{ik}g^{kl}b_{lj}\end{pmatrix}\;,
 \ee
which combines the space-time metric $g_{ij}$ and the Kalb-Ramond
two-form
$b_{ij}$. Here, $M,N,\ldots =1,\ldots, 2D$ are fundamental $O(D,D)$ indices,
where $D$ denotes the total number of space-time dimensions. Being an
element of $O(D,D)$, the generalized metric satisfies
 \be\label{constR1}
  {\cal H}^{MK}{\cal H}_{KN} \ = \ \delta^{M}{}_{N}\;,
 \ee
where
 \be\label{constR2}
  {\cal H}^{MN} \ \equiv  \
   \eta^{MK}\,\eta^{NL}\,{\cal H}_{KL}\;, \qquad
  \eta^{MN} \ = \ \begin{pmatrix}    0 & {\bf 1} \\[0.5ex]
  {\bf 1} & 0 \end{pmatrix}\;,
 \ee
and $\eta^{MN}$ is
the $O(D,D)$
invariant metric that will be used  to
raise and lower $O(D,D)$ indices.
The theory also includes the duality invariant dilaton field $d$
related to the standard dilaton $\phi$ via the field redefinition
$e^{-2d}=\sqrt{g}e^{-2\phi}$.

Double field theory features a
gauge symmetry parameterized
by an $O(D,D)$ vector parameter $\xi^{M}=(\tilde{\xi}_i\,,\xi^i)$ that combines
the diffeomorphism parameter $\xi^i$ and the $b$-field gauge parameter
$\tilde{\xi}_i$.
We will
refer to this gauge symmetry as `generalized diffeomorphisms'.
It acts on the fundamental variables as:
 \be\label{gendiff}
  \begin{split}
   \delta_{\xi}{\cal H}_{MN} \ &= \
   \xi^{P}\partial_{P}{\cal H}_{MN}+\big(\partial_{M}\xi^{P}-\partial^{P}\xi_{M}\big){\cal H}_{PN}
   +\big(\partial_{N}\xi^{P}-\partial^{P}\xi_{N}\big){\cal H}_{MP}\;, \\
   \delta_{\xi}\big(e^{-2d}\big) \ &= \ \partial_{M}\big(\xi^{M}e^{-2d}\big)\;,
  \end{split}
 \ee
where $\partial_{M}=(\tilde{\partial}^i,\partial_i)$ is the partial derivative
with respect to the doubled coordinates $X^M=(\tilde{x}_i,x^i)$.
We see that $e^{-2d}$ transforms as a scalar density.
The transformation rule in the top line of
(\ref{gendiff}) defines a generalized Lie derivative
$\delta_\xi {\cal H}_{MN} = \widehat {\cal L}_\xi  {\cal H}_{MN}$,
that can  be defined similarly for arbitrary $O(D,D)$ tensors.
An $O(D,D)$ tensor transforming under generalized diffeomorphisms with a generalized Lie derivative is called a generalized tensor.
The double field theory action
can be written as
 \be\label{DFTaction}
  S \ = \ \int dx d\tilde{x}\, e^{-2d}\,{\cal R}({\cal H},d)\;,
 \ee
where ${\cal R}$ is an $O(D,D)$ invariant function of ${\cal H}$
and $d$ that is a generalized scalar,
 \be\label{Rvar}
  \delta_{\xi}{\cal R}({\cal H},d) \ = \ \xi^{P}\partial_P{\cal R}({\cal H},d)\;,
 \ee
making the gauge invariance of (\ref{DFTaction}) manifest.
In order to verify the gauge variation (\ref{Rvar}) the following `strong
constraint' is required:
 \be\label{STRONG}
  \eta^{MN}\partial_{M}\partial_{N} \ = \ \partial^{M}\partial_{M} \ = \ 0\;,
   \ee
when acting on arbitrary fields and parameters and all their products.
This constraint implies that locally all fields depend only on half of
the coordinates, e.g., only the $x^i$.

The scalar ${\cal R}$ can be
viewed as
a generalized scalar curvature:  it reduces to the scalar curvature
when we set $b = \phi =0$ and choose the duality frame
$\tilde\partial =0$.
Moreover, the variation of (\ref{DFTaction}) with respect to ${\cal H}_{MN}$
gives rise to an $O(D,D)$ tensor
${\cal R}_{MN} ({\cal H}, d)$
 that is in fact
a generalized tensor and can be viewed
as a generalized Ricci tensor; its non-vanishing components reduce
 to the Ricci tensor
when we set $b = \phi =0$ and choose the duality frame
$\tilde\partial =0$.
Given this similarity with the corresponding tensors of Riemannian geometry it is natural to look for a systematic
way to construct these curvatures
starting with Christoffel-like connections and
a generalized
version of the Riemann tensor.
Indeed, it would be useful
to have a T-duality covariant generalization of the full Riemann
tensor in order to write general higher-derivative or $\alpha^{\prime}$
corrections to the effective action.

In searching for a generalized four-index Riemann tensor ${\cal R}_{MNPQ}$ it is useful to make
a list of properties that we may want this tensor to satisfy:
\begin{enumerate}

\item  It is a tensor under $O(D,D)$.

\item  It is  a tensor under generalized diffeomorphisms.

\item  It gives the generalized tensors ${\cal R}_{MN}$ and ${\cal R}$ upon suitable contractions.

\item  It is expressed in terms of the physical fields ${\cal H}_{MN}$
and~$d$.

\end{enumerate}
Property (1) ensures proper behavior under
T-duality and property (2)
ensures
proper behavior under gauge transformations.
Property (3) implies that, as in Riemannian geometry, the Riemann
tensor contains the information in Ricci and the information in the
scalar curvature.\footnote{The analogy with Riemannian geometry
is not complete:  there is no contraction of
${\cal R}_{MN}$ that gives~${\cal R}$.}
Property (4) means that the tensor is `physical', or fully determined.
We could also demand some additional properties that would
establish a close relation of ${\cal R}_{MNPQ}$ to the familiar
Riemann tensor. In analogy to the situation with ${\cal R}_{MN}$ and
${\cal R}$ we could demand that
\begin{enumerate}

\item [(A)]  For $b=\phi=0$ and $\tilde \partial=0$ some  components of
${\cal R}_{MNPQ}$ reduce  to the Riemann tensor.

\end{enumerate}
If property (4) holds, property (A) has a clear meaning.  If property (4) does not hold some components of ${\cal R}_{MNPQ}$
may be determined and some may not; we need only
study the former to test (A).

Some time ago Siegel developed
a vielbein formalism with a local $GL(D)\times GL(D)$ tangent
space symmetry~\cite{Siegel:1993th}. Introducing connections for this tangent space symmetry
he
defined invariant curvatures, but
not all
connections can be expressed in terms of the physical fields by
imposing covariant constraints. The scalar curvature and Ricci tensor
can be defined in a way that is independent of the undetermined connections,
but there does not appear to be an uncontracted Riemann tensor that
depends only on
physical fields.
Interestingly, in Batalin-Vilkovisky
quantization, a formalism based on antisymplectic geometry, a similar
phenomenon occurs: 
connections exist for which their undetermined components 
drop out  
of the curvature scalar~\cite{Batalin:2007xi}.

In this paper we will revisit these issues in a purely metric-like formalism.
We work solely with the generalized metric ${\cal H}_{MN}$
and the dilaton
and there are no additional gauge redundancies. This  
is equivalent to
Siegel's formulation and may 
be derived from it  by  imposing a vielbein postulate that relates the
Christoffel-like connections to the spin-connection~\cite{Hohm:2010xe}.
This will be briefly explained in the appendix. 
We find it simpler and more illuminating, however,
 to present the metric-like formalism in a self-contained
fashion.
A closely related formulation has been developed before
in  useful papers
by Jeon, Lee, and Park~\cite{Jeon:2010rw,Jeon:2011cn}.
Many of our results have a direct analogue in the frame formalism 
of Siegel and some have appeared in~\cite{Jeon:2010rw,Jeon:2011cn}.  
Finally, generalized geometry \cite{arXiv:0710.2719}
also features closely related
connections and curvatures; see
\cite{Coimbra:2011nw} for a recent concise exposition.

We investigate systematically within the formalism
if  there is a  ${\cal R}_{MNPQ}$
that satisfies the properties listed above ((1) through (4), and (A)).
Our investigation confirms the
existence of a duality covariant generalized Riemann tensor that
determines ${\cal R}_{MN}$ and ${\cal R}$.  Thus properties (1), (2),
and (3) hold.
We find, however, that ${\cal R}_{MNPQ}$
is not fully determined in terms of the physical fields: property (4)
does not hold.
We show  that
this is a simple consequence of an algebraic Bianchi identity of the Riemann tensor.  In fact, property (A) does not hold either: the components of ${\cal R}_{MNPQ}$ that do not contain undetermined connections
are zero.

The generalized metric formulation
differs from
Riemannian geometry in that the metric is a constrained object;
it satisfies (\ref{constR1})--(\ref{constR2}). As a consequence, there are projectors
 \be\label{projIntro}
  P_{M}{}^{N} \ = \ \frac{1}{2}\big(\delta_{M}{}^{N}-{\cal H}_{M}{}^{N}\big)\;, \qquad
  \bar{P}_{M}{}^{N} \ = \ \frac{1}{2}\big(\delta_{M}{}^{N}+{\cal H}_{M}{}^{N}\big)\;,
 \ee
satisfying $P + \bar P = 1$,
$P\bar{P}= 0$, $P^2 =P$ and $\bar{P}^2 = \bar{P}$.
They allow us to project onto a `left-handed' or `right-handed' subspace.
This is the
analogue of the factorized
tangent space
group $GL(D)\times GL(D)$ in the frame formulation, and equivalence
of the two formalisms then requires the projectors to be
covariantly constant.
Jeon, Lee and Park~\cite{Jeon:2010rw,Jeon:2011cn}
postulate an expression for the Christoffel
symbols in terms of the physical fields that
satisfies
this condition.
The resulting `covariant derivatives', however,
do not transform covariantly in general, but only for certain projections
and contractions.   The reason is that the imposition
of covariant constraints only determines part of the connections, and their
ansatz effectively sets the undetermined connections to zero, thereby
violating covariance. Here we
follow a somewhat
different route.  As in the frame formalism,
we work with proper connections and fully covariant expressions
by keeping those connection components that are not determined by the physical fields.
For the final results on Ricci and scalar curvature tensors
for which the undetermined connections drop out,
our results are in full agreement with the most recent
work \cite{Jeon:2011cn}.
We also establish
differential Bianchi identities that 
have not appeared before in such a metric-like formalism.

An important motivation for this work was the construction of higher-derivative
or $\alpha^{\prime}$ corrections involving the full Riemann tensor. Thus,
in the second part of this paper we
ask if there is a
manifestly $O(D,D)$ invariant function of the
generalized metric (\ref{firstH}),
quartic in derivatives,
that reduces
to the square of the Riemann tensor in some
T-duality frame. In fact, even if there is no physical
 ${\cal R}_{MNPQ}$,
 one can imagine
an expression
that reproduces the square of the
Riemann tensor,
but is not
the square of an $O(D,D)$ tensor.
We find, however, that for general $D$ such a construction is impossible, showing
that generic $\alpha^{\prime}$ corrections
cannot be written in terms of the generalized metric  defined in
terms of $g$ and $b$
as in (\ref{firstH}).

In hindsight, this result is not too surprising in view of similar
results obtained for dimensionally reduced theories. It has
been shown by Meissner that $\alpha^{\prime}$-corrected  supergravity,
reduced to one dimension,
can be written in a
T-duality invariant way
if the formula for the generalized metric in terms of the
physical fields
receives $\alpha^{\prime}$
corrections~\cite{Meissner:1996sa}.
We
discuss in the conclusions the possible  implications
of this fact for our analysis.

\section{Christoffel connections and invariant curvatures}

In this section we introduce Christoffel-type connections and determine their
transformation behavior by requiring that they give rise to derivatives
that are covariant under generalized diffeomorphisms. In terms of these connections
we define an $O(D,D)$ covariant Riemann tensor that is also a generalized
tensor. Next, we impose covariant constraints on the connections in
order to express them in terms of the physical fields. It turns out that
this leaves undetermined components, which we analyze systematically.

\subsection{Connections and curvatures}\setcounter{equation}{0}

$O(D,D)$ tensors are said to be generalized
tensors if they transform with generalized Lie derivatives
under
generalized diffeomorphisms parametrized by $\xi^{M}$.
The generalized Lie derivative is defined on
generalized
vectors as
 \be
 \begin{split}
  \delta_{\xi}A^{M} \ &= \ \widehat{\cal L}_{\xi}A^{M} \ \equiv \ \xi^{N}\partial_{N}A^{M}+\big(\partial^{M}\xi_{N}-\partial_{N}\xi^{M}\big)A^{N}\;, \\
   \delta_{\xi}A_{M} \ &= \ \widehat{\cal L}_{\xi}A_{M} \ \equiv \ \xi^{N}\partial_{N}A_{M}+\big(\partial_{M}\xi^{N}-\partial^{N}\xi_{M}\big)A_{N}\;,
 \end{split}
 \ee
and is defined
similarly on tensors with an arbitrary number of upper and lower $O(D,D)$ indices.
For a generalized scalar $S$ the generalized Lie derivative is just given by
the transport term.  The  partial derivative of a scalar is a generalized
vector since
 \bea
  \delta_{\xi}\left(\partial_{M}S\right) \ = \
  \partial_{M}\left(\xi^{P}\partial_{P}S\right) \ = \
  \xi^{P}\partial_{P}(\partial_{M}S)+\partial_{M}\xi^{P}\,\partial_{P}S
  -\partial^{P}\xi_{M}\,\partial_{P}S\;,
 \eea
where we are allowed to add the last term because it vanishes by the constraint (\ref{STRONG}).
Next, we define a covariant derivative of a vector
by introducing a connection $\Gamma$:
 \be
 \begin{split}
  \nabla_{M}A_{N} \ & \equiv \
   \ \partial_{M}A_{N}-\Gamma_{MN}{}^{K}A_{K}\;,\\
   \nabla_{M}A^{N} \ & \equiv \
   \ \partial_{M}A^{N}
   +\Gamma_{MK}{}^{N}A^{K}\,.
 \end{split}
 \ee
The transformation property of the connection is determined by
the condition that the above derivatives be generalized tensors.
A short calculation shows that one must have
 \bea\label{conntrans}
  \delta_{\xi}\Gamma_{MN}{}^{P} \ = \ \widehat{\cal L}_{\xi}\Gamma_{MN}{}^{P}
  +\partial_{M}\partial_{N}\xi^{P}-\partial_{M}\partial^{P}\xi_{N}\;.
 \eea
The first two terms on the right-hand side
are familiar
and the last one
is due to the extra terms in
the generalized Lie derivative. That  last term
implies
that the connection cannot be
chosen to be symmetric in its first two indices $M$ and $N$.
We will
let $\Delta_\xi$ denote all non-covariant terms in a trasformation law:  $\delta_\xi W \ = \ \widehat{{\cal L }}_\xi W + \Delta_\xi W$, for any
$O(D,D)$ tensor $W$.
We then have
 \be\label{DelGam}
  \Delta_{\xi}\Gamma_{MNK} \ = \ 2\partial_{M}\partial_{[N}\xi_{K]}\;,
 \ee
where, as usual, we raise and lower all indices with $\eta$.

Given these connections
we can define curvature and torsion
through the commutator of covariant derivatives,
 \bea\label{commutator}
  \big[\nabla_{M},\,\nabla_{N}\big]A_{K} \ = \ -R_{MNK}{}^{L}\,A_{L}
  -T_{MN}{}^{L}\,\nabla_{L}A_{K}\;.
 \eea
One finds
 \bea\label{conRieTor}
 \begin{split}
  R_{MNK}{}^{L} \ &= \ \partial_{M}\Gamma_{NK}{}^{L}-\partial_{N}\Gamma_{MK}{}^{L}
  +\Gamma_{MQ}{}^{L}\Gamma_{NK}{}^{Q}-\Gamma_{NQ}{}^{L}\Gamma_{MK}{}^{Q}\;, \\[0.8ex]
  T_{MN}{}^{K} \ &= \ 2\,\Gamma_{[MN]}{}^{K}\;.
 \end{split}
 \eea
By definition $R$ is antisymmetric on the first two indices,
\be
R_{MNK}{}^{L}  \ = \  -  R_{NMK}{}^{L}\,.
\ee
There will also be an antisymmetry in the last two indices after
the imposition of constraints.  Lowering the $L$ index in $R_{MNK}{}^L$
we have
 \be
 \label{Rie-alldown}
  R_{MNKL} \ = \ \partial_{M}\Gamma_{NKL}
  -\partial_{N}\Gamma_{MKL}
  +\Gamma_{MQL}\Gamma_{NK}{}^{Q}-\Gamma_{NQL}\Gamma_{MK}{}^{Q}\;.
   \ee

It turns out that neither $R$ nor $T$ is a generalized tensor.
The non-covariant transformation of the torsion tensor follows
directly by applying (\ref{DelGam}) to the definition in (\ref{conRieTor}).
The non-covariant transformation of $R$ follows by a slightly longer
but straightforward computation. In total, one finds
 \be\label{noncov}
\begin{split}
  \Delta_{\xi}R_{MNK}{}^{L} \ &= \
  -2\, \partial^P\partial_{[M}\xi_{N]}\,\Gamma_{PK}{}^{L}\;, \\[0.8ex]
   \Delta_{\xi}T_{MN}{}^{L} \ &= \ - 2\, \partial^L \partial_{[M}\xi_{N]}\;.
 \end{split}
 \ee
While each of the two terms on the right-hand side of (\ref{commutator})
fails to transform covariantly the sum must since the left-hand side is
manifestly covariant.  This can be readily checked; acting with
$\Delta_\xi$ on the right-hand side of (\ref{commutator}) gives
\be
\begin{split}
 -\Delta_\xi R_{MNK}{}^{L}\,A_{L}
  -\Delta_\xi T_{MN}{}^{L}\,\nabla_{L}A_{K}
   \ & = \   2\,\partial_{[M}\partial^{P}\xi_{N]}\,\Gamma_{PK}{}^{L}\,A_{L}
  + 2\partial_{[M}\partial^{P}\xi_{N]}\,(\partial_{P}A_{K}
  - \Gamma_{PK}{}^LA_L ) \,, \nonumber
\end{split}
\ee
where use was made of (\ref{noncov}).  The term with a $\partial_PA_K$
vanishes by the strong constraint and the other two terms cancel each other so that, as expected,
\be
\Delta_\xi \bigl( -R_{MNK}{}^{L}\,A_{L}
  - T_{MN}{}^{L}\,\nabla_{L}A_{K} \bigr) \ = \ 0 \,.
\ee

Although $R_{MNKL}$  is not a generalized tensor it can be made so by the simple
following modification.  Note that  the first equation
 in~(\ref{noncov}) can
be written as
\be
  \Delta_{\xi}R_{MNKL} \ = \ -2\, \partial_Q\partial_{[M}\xi_{N]}\,
  \Gamma^Q{}_{KL}  \ = \
  -\, (\Delta_\xi \Gamma_{QMN} )\,\Gamma^Q{}_{KL}\,,
\ee
by use of (\ref{DelGam}).  This equation makes it easy to see
 that  ${\cal R}_{MNKL}$, defined by
 \be\label{covRiem}
  {\cal R}_{MNKL} \ \equiv  \ R_{MNKL}+R_{KLMN}+\Gamma_{QMN}\Gamma^{Q}{}_{KL}\;,~~
 \ee
is a generalized tensor.  By definition ${\cal R}$ is symmetric under
the interchange of the first and second pair of indices:
\be
\label{exchsym}
{\cal R}_{MNKL}  \ = \    {\cal R}_{KLMN} \,.
\ee
The antisymmetry $R_{MNKL} = - R_{NMKL}$ in the first pair of
indices
does not immediately
carry over to ${\cal R}_{MNKL}$ but it will after the imposition of
constraints on the connection.

\subsection{Constraints on the connection}

We now impose four constraints in order to determine part of the connections in terms of the
physical fields ${\cal H}$ and $d$. 
These constraints follow from the constraints 
of Siegel's frame formalism given in \cite{Siegel:1993th}, as will be reviewed in the appendix, 
and are also satisfied by the connection-like objects postulated in \cite{Jeon:2011cn}.   
The first two set some components of the connection equal to zero and
do not involve ${\cal H}$ 
or $d$. The third constraint involves ${\cal H}$
and the fourth involves the dilaton $d$.   As we will see in the following
section, the connection is not fully determined by these four constraints.

 \begin{itemize}
  \item[(1)] Covariant constancy of $\eta_{MN}$:
   \be
   \label{constraint1}
    \nabla_{M}\eta_{NP}
    \ = \ \partial_M \eta_{NP} -  \Gamma_{MN}{}^Q \eta_{QP}
    - \Gamma_{MP}{}^Q\eta_{NQ} \ =  \ 0\;\quad\Rightarrow \quad \;   \Gamma_{MNP} + \Gamma_{MPN} \ = \ 0 \;,
    \ee
   where we recall that $\eta$ is a constant matrix and that indices are lowered with $\eta$.
  This equation means that the connection is antisymmetric in the
  last two indices,
  \be
  \Gamma_{MNP} \ = \ - \Gamma_{MPN} \,.
  \ee

    \item[(2)]
   Generalized torsion constraint:  We
   demand
   that
   the generalized Lie derivative of a vector can be evaluated with
   an identically looking formula where partial derivatives are replaced
   by covariant derivatives,
    \be\label{GenTorsion}
     \widehat{\cal L}_{\xi}V_{M} \ \equiv  \  \xi^N \partial_N V_M
     + 2\partial_{[M}\xi_{N]}\,V^{N}  \ = \  \xi^{N}\nabla_{N}V_{M}+2\nabla_{[M}\xi_{N]}\,V^{N}\ = \  \widehat{\cal L}_{\xi}^{\;\nabla} V_M\,.
     \ee
 Here $\widehat{\cal L}_{\xi}^{\;\nabla}$ denote the generalized Lie derivative with $\partial$
      replaced by $\nabla$.     Put differently,
      we are setting to zero a generalized torsion tensor ${\cal T}$
      defined by \cite{Coimbra:2011nw}
      \be\label{torDef}
        \big( \widehat{\cal L}_{\xi}^{\;\nabla}- \widehat{\cal L}_{\xi}\big)V_{M} \ = \  {\cal T}_{MNK}\xi^{N}V^{K} \;.
      \ee
        A short calculation gives \cite{Jeon:2010rw}
     \be\label{gtcnstr1}
   {\cal T}_{MNK}  \ = \ \Gamma_{MNK} -  \Gamma_{NMK} +  \Gamma_{KMN}  \ = \ T_{MNK}+\Gamma_{KMN}\,.
   \ee
 As defined in (\ref{torDef}) ${\cal T}_{MNK}$ is manifestly
a generalized tensor,
and
this can also be checked directly with (\ref{DelGam}).
Our constraint sets this generalized torsion to zero:
\be
{\cal T}_{MNK}  \ = \ \Gamma_{MNK} -  \Gamma_{NMK} +  \Gamma_{KMN}
\ = \ 0 \,.\ee
Using constraint (1) we find that the sum of cyclic index permutations vanishes:
   \be
   \label{cyclicconn}
   \Gamma_{MNK} +  \Gamma_{NKM} +  \Gamma_{KMN} \ = \ 0 \,.
   \ee
This property, given constraint (1), is equivalent to the condition that
the totally antisymmetric part of the connection vanishes:
     \be
  \Gamma_{[MNK]} \ = \ 0 \; .
    \ee

   \item[(3)] Covariant constancy of ${\cal H}_{MN}$:
    \be
    \label{constraint3}
     \nabla_{M}{\cal H}_{NK} \ = \ \partial_{M}{\cal H}_{NK}-\Gamma_{MN}{}^{P}{\cal H}_{PK}
     -\Gamma_{MK}{}^{P}{\cal H}_{NP} \ = \ 0\;.
    \ee

   \item[(4)] Partial integration in presence of dilaton density:
    \be\label{dilatonconstr}
     \int e^{-2d}\,V\nabla_{M}V^{M} \ = \ -\int e^{-2d}\,V^{M}\nabla_{M}
     V\;.
     \ee
     This condition results in
     \be\label{Gammatraced}
     \; \Gamma_{M} \ \equiv \ \Gamma_{NM}{}^{N} \ = \ -2\partial_{M}d\;.
    \ee
    Equivalently, this condition means that the
    covariant divergence of a vector is computed using the density
    $e^{-2d}$:
    \be\label{covdiv}
  \nabla_{M}V^{M} =  \partial_M A^M  + \Gamma_{MK}{}^M A^K  \  = \
  e^{2d} \partial_M ( e^{-2d} A^M) \,.
  \ee

\end{itemize}

\subsection{Solving the constraints}

\subsubsection{The first constraint}

  We can derive a number of conclusions from  constraint
  (\ref{constraint1}) that
  states the covariant constancy of the $O(D,D)$
  metric $\eta_{MN}$.  This constraint makes the connection antisymmetric on its last two indices.
Now
consider  the curvature $R_{MNKL}$ in (\ref{Rie-alldown}). Using the antisymmetry condition, the last two terms are rewritten as
 \be
 \label{Rie-alldown-final}
  R_{MNKL} \ = \ \partial_{M}\Gamma_{NKL}
  -\partial_{N}\Gamma_{MKL}
  -\Gamma_{MLQ}\Gamma_{NK}{}^{Q}
  +\Gamma_{MKQ}\Gamma_{NL}{}^{Q}\;,
   \ee
making it clear that $R_{MNKL}$ is now also antisymmetric in the
last two indices.
Since it is also antisymmetric
 in its first  two indices we have in total
 \be
 \label{Rtrans}
  R_{MNKL} \ = \ -R_{NMKL} \ = \ -R_{MNLK} \,.
   \ee
Still, there is no simple relation between $R_{MNKL}$ and $R_{KLMN}$.
It also follows from the above and (\ref{covRiem}) that  ${\cal R}$ shares
those same symmetries,
 \be
 \label{calRtrans}
  {\cal R}_{MNKL} \ = \ -{\cal R}_{NMKL} \ = \ -{\cal R}_{MNLK}
  \;.~~
 \ee
Together with (\ref{exchsym}) we see that ${\cal R}_{MNKL}$ satisfies
the familiar properties of the Riemann tensor.  One missing property, the algebraic
Bianchi identity, will follow after the imposition of the second constraint.

\subsubsection{The second constraint}

Let us now see what conclusions follow from the
vanishing of the generalized torsion.
First, we note that the formula for
the torsion in terms of the connection can be simplified.
With (\ref{gtcnstr1}) it follows from ${\cal T}_{MNK}=0$ that
 \be
  T_{MNK} \ = \ -\Gamma_{KMN}\;.
 \ee
An important consequence of the first two constraints is that we
have the Bianchi identity
 \be\label{algebraicBianchi}
 {\cal R}_{[MNK]L} \ = \ 0\;,
 \ee
as also noted in~\cite{Jeon:2011cn}.
Given the symmetries (\ref{calRtrans}), this is equivalent to
\be
{\cal R}_{MNKL} + {\cal R}_{NKML} + {\cal R}_{KMNL} \ = \ 0 \,.
\ee
In Riemannian geometry this formula follows directly from the expression
for the Riemann tensor in terms of a torsion-less connection.  In the present case the equation requires
\be
\begin{split}
&{R}_{MNKL} +  R_{NKML} + { R}_{KMNL} \\
+\,& { R}_{KLMN} + { R}_{MLNK} + {R}_{NLKM}  \\
+\,& \Gamma_{QMN} \Gamma^Q{}_{KL}
+ \Gamma_{QNK} \Gamma^Q{}_{ML}
+\Gamma_{QKM} \Gamma^Q{}_{NL}    \ = \ 0 \,. \end{split}
\ee
This equation is readily verified using (\ref{conRieTor}):  there are twelve
terms of the form $\partial \Gamma$ that combine into four groups of
three terms that vanish separately,  there are fifteen $\Gamma\Gamma$
terms that combine into three groups of five terms that vanish separately.

\medskip
Finally, we derive a formula  for the exact variation of
${\cal R}_{MNKL}$ upon a finite variation  $\Gamma \to \Gamma
+ \delta \Gamma$ of the connection.  Beginning with (\ref{Rie-alldown}), a short
calculation gives
 \be
 \label{varR}
 \begin{split}
  R_{MNKL} (\Gamma + \delta \Gamma) \ =&~ \ {R}_{MNKL}(\Gamma)
  \,+\, 2\,{\nabla}_{[M}\delta\Gamma_{N]KL} \\
&~  +2{\Gamma}_{[MN]}{}^{P}
  \delta\Gamma_{PKL}+2\,\delta \Gamma_{[M|QL|}
 \, \delta\Gamma_{N]K}{}^{Q} \;.
 \end{split}
 \ee
In obtaining the above we only had to use the antisymmetry of the connection in the last two indices (constraint 1).  The covariant
derivatives on the above right-hand side use $\Gamma$.
To obtain the analogous result for ${\cal R}_{MNKL}$ we use
the above and (\ref{covRiem}).  This time a short calculation
gives
 \be\label{varcalR}
  \begin{split}
   {\cal R}_{MNKL} (\Gamma + \delta \Gamma) \ = \  \ & {\cal R}_{MNKL} (\Gamma) \ +\ 2\,{\nabla}_{[M}\delta\Gamma_{N]KL} \ + \ 2\,{\nabla}_{[K}
   \delta\Gamma_{L]MN}
   \\[1.0ex]
   &\hskip-8pt +2\,\delta\Gamma_{[M|QL|}\,\delta\Gamma_{N]K}{}^{Q}
   \, +\, 2\,\delta\Gamma_{[K|QN|}\,\delta\Gamma_{L]M}{}^{Q}
   \,+\, \delta\Gamma_{QMN}\,\delta\Gamma^{Q}{}_{KL}\;.
  \end{split}
 \ee
In deriving this result we had to use the second constraint in the form
(\ref{cyclicconn}). Note that the terms of the form $\Gamma \delta \Gamma$ in $R(\Gamma + \delta \Gamma)$ cancel out in
${\cal R}(\Gamma + \delta \Gamma)$.

\subsubsection{The third constraint}

The constraint (\ref{constraint3})
demands the covariant constancy of the generalized
metric. To explore immediate consequences of this additional constraint
consider the projectors (\ref{projIntro})
  \be
  P_{M}{}^{N} \ = \ \frac{1}{2}\big(\delta_{M}{}^{N}-{\cal H}_{M}{}^{N}\big)\;, \qquad
  \bar{P}_{M}{}^{N} \ = \ \frac{1}{2}\big(\delta_{M}{}^{N}+{\cal H}_{M}{}^{N}\big)\;,
 \ee
which satisfy
 \be\label{projiden}
  P\,\bar{P} \ = \ 0\;, \qquad P^2 \ = \ P\;, \qquad \bar{P}^2 \ = \ \bar{P}\;.
 \ee
Since $\eta$
is covariantly constant by constraint (1) and
${\cal H}$ is covariantly constant by constraint (3),
the projectors are also covariantly
constant:
\be
\nabla_K  P_{M}{}^{N}  \ = \   \nabla_K  \bar{P}_{M}{}^{N} \ = \ 0 \,.
\ee
We now discuss how to use this result to solve completely
the constraint.
For this purpose we will introduce a notation for indices that are projected.
We will have two kinds of indices:
barred, with a dash on top, and un-barred, or more properly, under-barred, with a dash below.
The index type depends on the projector that is used to obtain it from
the un-projected index.  The barred index is associated with the $\bar P$ projector and
the under-barred index is associated with the $P$ projector. Thus, we will have
\be
\label{defprojindices}
\begin{split}
W_{\nin{M}} \ &\equiv  \  P_M{}^N \, W_N \,,\\
W_{\bar{M}} \ &\equiv \  \bar P_M{}^N \, W_N\,.
\end{split}
\ee
Note that this implies that
\be
\label{sumtwo}
W_M \ = \ W_{\nin{M}}  \ + \  W_{\bar M} \,.
\ee
We raise or lower projected indices
with the metric $\eta$:
\be
\label{defprojindices-raised}
\begin{split}
W^{\nin{M}} \ \equiv  \ \eta^{MQ} W_{\nin{Q}}  \ &= \  \eta^{MQ} \,P_Q{}^N \, W_N  \ = \  P^{MN} W_N  \,, \\
W^{\bar{M}} \ \equiv \  \eta^{MQ} W_{\bar{Q}}  \ &= \  \eta^{MQ} \,\bar{P}_Q{}^N \, W_N  \ = \  \bar{P}^{MN} W_N  \,,
\end{split}
\ee
so that one can simply use the projector with indices up or down to
define a projected index. Contraction of projected indices of different
types vanish. For example,
\be
W^{\nin{M}}  Y_{\bar{M}}  \ = \ 0\,.
\ee
Contraction of like-wise
projected indices can be done with a single projector:
\be
\begin{split}
\label{contract-project}
W^{\bar{M}} Y_{\bar{M}} \ &= \  \bar P^{MQ}  \bar P_{M}{}^R  \,W_Q Y_R
 \ = \  \bar P^{QR} W_Q Y_R\,, \\[0.5ex]
 W^{\nin{M}} Y_{\nin{M}} \ &= \  P^{MQ}   P_{M}{}^R  \,W_Q Y_R
 \ = \   P^{QR} W_Q Y_R\,.
 \end{split}
\ee
A contraction of unprojected indices can be written as a sum of contractions of like-wise
projected indices.  Indeed,
\be
W^M Y_M  \ = \  (W^{\nin{M}} + W^{\bar M}) ( Y_{\nin{M}} + Y_{\bar{M}})
 \ = \  W^{\nin{M}} Y_{\nin{M}}  \ + \ W^{\bar M} Y_{\bar{M}} \,.
\ee
We will occasionally use tensors with mixed indices. So for example,
we could have an object
\be
W_{M \,\nin{N} \,K}  \ = \  P_{N}{}^Q \, W_{M QK}\,.
\ee
There is no possible confusion: an index without a bar or under-bar is
unprojected.
As a final remark on the use of these indices we note that
in any tensor equality with a number of free unprojected
 indices (appearing both on the
left-hand side and the right-hand side) we can simply replace any
unprojected index by like-wise
projected indices on both sides of the equality.
Thus, for example,  $ W_{MN}  =  Y_{MN}$ implies
$W_{\nin{M}\bar{N}}  =  Y_{\nin{M}\bar{N}}$, as well as several other
equalities.

When dealing with objects with projected indices, we will say that
the object is of type $(k,l)$ if it has $k$ under-barred indices and
$l$ barred indices. Thus, for example, given an $O(D,D)$ tensor
$A_{MNP}$ we have
\be
\label{type-defined}
\hbox{Type } (3,0): ~~  A_{\,\nin{M}\nin{N}\nin{P}} \,, ~~~~~ \hbox{Type }(2,1):~~ A_{\,\bar{M}\,\nin{N}\nin{P}}\;, ~A_{\,\nin{M} \bar{N}\nin{P}} \;, ~
A_{\,\nin{M} \nin{N} \bar{P}} \;,  ~\hbox{etc.}
\ee

\medskip
Let us now consider the connection $\Gamma_{MNK}$.  By repeated
use of (\ref{sumtwo}) on each index we have
\be\label{explrepr99}
 \begin{split}
 \Gamma_{MNK}   \ &= \ \;
 \Gamma_{\,\nin{M}\,\nin{N}\,\nin{K}} + \Gamma_{\nin{M}\,\nin{N}\bar K}
+   \Gamma_{\nin{M}\bar N\,\nin{K}}    +\Gamma_{\nin{M}\bar N\bar K}
    \\[1.0ex]
&~~+\Gamma_{\bar M \,\nin{N}\,\nin{K}} +\Gamma_{\bar M\,\nin{N}\bar K}
+\Gamma_{\bar M\bar N\,\nin{K}}
+\Gamma_{\bar M\bar N\bar K}\,.
 \end{split}
 \ee
From the comments above it follows that the symmetries of $\Gamma$ arising from the first two constraints carry over to
the projected $\Gamma$.
Thus, for example, $\Gamma_{\,\nin{M}\,\nin{N}\bar K}  =  - \Gamma_{\,\nin{M}\bar K \,\nin{N}}$.
The cyclicity condition
on the three indices also holds for any choice of index type.

Using the symmetry conditions on
$\Gamma$ we can rewrite (\ref{explrepr99}) as follows:
\be\label{explrepr9999}
 \begin{split}
 \Gamma_{MNK}   \ &= \ \;
 \Gamma_{\nin{M}\nin{N}\nin{K}} + \Gamma_{\nin{M}\nin{N}\bar K}
-   \Gamma_{\nin{M}\nin{K}\bar N}
- (-\Gamma_{\bar N\nin{M}\bar K} + \Gamma_{\bar K \nin{M} \bar N})
    \\[1.0ex]
&~~-(\Gamma_{\nin{N}\nin{K}\bar M} - \Gamma_{\nin{K}\nin{N}\bar M })
+\Gamma_{\bar M\nin{N}\bar K}
-\Gamma_{\bar M\nin{K}\bar N}
+\Gamma_{\bar M\bar N\bar K}\,.
 \end{split}
 \ee
We then regroup the terms to find
\be\label{explrepr99998}
 \begin{split}
 \Gamma_{MNK}   \ &= \ \;
 \Gamma_{\nin{M}\nin{N}\nin{K}}  +\Gamma_{\bar M\bar N\bar K} \\[1.0ex]
&~~
+ \Gamma_{\nin{M}\nin{N}\bar K}
-   \Gamma_{\nin{M}\nin{K}\bar N}
-\Gamma_{\nin{N}\nin{K}\bar M} + \Gamma_{\nin{K}\nin{N}\bar M }
    \\[1.0ex]
&~~
+\Gamma_{\bar M\nin{N}\bar K}
-\Gamma_{\bar M\nin{K}\bar N}  - \Gamma_{\bar K \nin{M} \bar N}
+\Gamma_{\bar N\nin{M}\bar K}\,.
 \end{split}
 \ee
This shows that  there are
just four structures that need to be determined:
\be
 \Gamma_{\,\nin{M}\,\nin{N}\,\nin{K}}  \,, ~~\Gamma_{\bar M\bar N\bar K}\,,
 ~~ \Gamma_{\nin{M}\,\nin{N}\bar K}\,,~
 \;\hbox{and}\;~~ \Gamma_{\bar M\,\nin{N}\bar K} \,.
\ee
As we will now see, the covariant constancy of the projector
determines the last two of these and leaves the first two undetermined.
Indeed, consider the equation
 \be\label{covconpro}
  \nabla_{M}P_{K}{}^{L} \ = \ \partial_{M}P_{K}{}^{L}-\Gamma_{MK}{}^{Q}P_{Q}{}^{L}+\Gamma_{MQ}{}^{L}P_{K}{}^{Q} \ = \ 0\;.
 \ee
We write this as
\be
\label{vmbw}
\partial_{M}P_{KL}+ P_L{}^{Q}\Gamma_{MQK}+
P_{K}{}^{Q}\Gamma_{MQL} \ = \ 0\,.
\ee
Multiplying by $\bar P_N{}^K$ the last term drops out and we get
\be\label{someprojection99}
 P_L{}^{Q} \bar P_N{}^K\Gamma_{MQK}
 \ = \ -\bar P_N{}^K\partial_{M}P_{KL} \ =
  \ -(\bar P \partial_M P)_{NL}\,,
\ee
or, equivalently,
\be\label{someprojection}
 \Gamma_{M\nin{L}\bar N}
\ = \ -(\bar P \partial_M P)_{NL}\,.
\ee
Acting with an additional projector we obtain,
\be\label{SomeComp}
\begin{split}
 \Gamma_{\nin{R}\,\nin{L} \bar N}
 \ &= \ -  P_R{}^M(\bar P \partial_M P)_{NL}\,,\\
 \Gamma_{\bar R\,\nin{L}\bar N}
 \ &= \ - \bar P_R{}^M(\bar P \partial_M P)_{NL}\,.
\end{split}
\ee
This determined the advertised components.  The
totally under-barred
component $\Gamma_{\nin{M}\nin{N}\nin{K}}$ of $\Gamma_{MNK}$
is not determined because it drops out of (\ref{vmbw}).  Indeed note
that
\be
P_L{}^{Q}\Gamma_{\nin{M}\nin{Q}\,\nin{K}}+
P_{K}{}^{Q}\Gamma_{\nin{M}\nin{Q}\nin{L}} \ = \ \Gamma_{\nin{M}\nin{L}\,\nin{K}}+
\Gamma_{\nin{M}\nin{K}\,\nin{L}} \ = \ 0 \,,
\ee
because of antisymmetry on the last two indices.  Of course, the
totally barred components  $\Gamma_{\bar M \bar N \bar K}$ are
also not determined.

\subsubsection{The fourth constraint}

This constraint determines the trace of the connection:
\be
\Gamma_N \ \equiv \ \Gamma_{MNK}\eta^{MK} \ = \ - 2\, \partial_N d\,.
\ee
To begin the analysis we compute the left-hand side of this relation using (\ref{explrepr99998}).
We get
 \be\label{explrepr99798}
\Gamma_N   \ = \ \;
\eta^{MK} \Gamma_{\nin{M}\nin{N}\nin{K}}  +\eta^{MK}
\Gamma_{\bar M\bar N\bar K} -  \eta^{MK} \Gamma_{\nin{M}\,\nin{K}\bar N}
+\eta^{MK}\Gamma_{\bar M\,\nin{N}\bar K}\,,
 \ee
where we noted that contractions of $\eta$ with $\Gamma$ are nonzero
only if the two indices to be contracted in the projected $\Gamma$ are of the same type.
Moving the undetermined components to the left-hand side
and recalling (\ref{sumtwo})
we obtain
  \be\label{exppr996798}
\eta^{MK} \Gamma_{\nin{M}\nin{N}\nin{K}}  +\eta^{MK}
\Gamma_{\bar M\bar N\bar K}
 \ = \   \Gamma_{\nin{N}}-\eta^{MK}\Gamma_{\bar M\,\nin{N}\bar K}
 \  + \  \Gamma_{\bar N}- \eta^{MK} \Gamma_{\nin{M}\bar N\,\nin{K}}   \,.
 \ee
From the above we obtain two equations for the two undetermined components, according to the type of $N$ index:
  \be
  \label{fortraceeqns}
  \begin{split}
  \eta^{MK} \Gamma_{\nin{M}\nin{N}\nin{K}}
 \ &= \   \Gamma_{\nin{N}}-\eta^{MK}\Gamma_{\bar M\,\nin{N}\bar K}  \ \equiv \ \phi_{\nin{N}}\,.
\\[0.4ex]
 \eta^{MK}\Gamma_{\bar M\bar N\bar K}
 \ &= \    \Gamma_{\bar N}- \eta^{MK} \Gamma_{\nin{M}\bar N\,\nin{K}}  \ \equiv \ \phi_{\bar N} \,.
  \end{split}
  \ee
Note that $\phi_{\nin{N}}$ and $\phi_{\bar N}$ are projected objects.  It is useful
to show that they arise from a single object $\phi_N$.
 This is what we do now.
We begin with
$\phi_{\nin{N}}$ and use (\ref{SomeComp}):
\be
\begin{split}
\phi_{\nin{N}} \ = & \,\  P_N{}^R \Gamma_R \ + \ \eta^{MK} \bar P_M{}^Q ( \bar P \partial_Q P)_{KN}\\
= & \, \  P_N{}^R \Gamma_R  \ - \   \bar P^{KQ} ( (\partial_Q\bar P)  P)_{KN}\\
= & \, \  P_N{}^R \Gamma_R  \ - \   \bar P^{KQ} \,(\partial_Q\bar P_{KR} )P_N{}^R\\
= & \, \  P_N{}^R \Bigl( \Gamma_R  \ - \   \bar P_{QK} \,\partial^Q
\bar P^K{}_R  \Bigr)\ = \  \  P_N{}^R \Bigl( \Gamma_R  \ - \
(\bar P \,\partial^Q \bar P)_{QR}  \Bigr)\,.
\end{split}
\ee
We note that in the final expression
 the reversed index combination
$(\bar P \partial^Q \bar P)_{RQ}$ would give zero contribution due to
the $P_N{}^R$ projector.  We can thus write,
\be
\phi_{\nin{N}} \ =    \  P_N{}^R \Bigl( \Gamma_R  \ - \
2 (\bar P \,\partial^Q \bar P)_{[QR]}  \Bigr)\,.
\ee
A completely analogous calculation gives
\be
\phi_{\bar N}  \ =    \  \bar P_N{}^R \Bigl( \Gamma_R  \ - \
2 ( P \,\partial^Q  P)_{[QR]}  \Bigr)\,.
\ee
We can easily verify that the terms in parenthesis in the two equations
above are equal.  Indeed,
\be
P\partial^Q P = -(1-\bar P)\partial^Q\bar P
 =  -\partial^Q \bar P +  \bar P\partial^Q \bar P \quad \to \quad
 (P\partial^Q P)_{[QR]} \ = \ (\bar P\partial^Q \bar P)_{[QR]}\,.
\ee
We can therefore write
\be
\label{singleff}
\phi_{\nin{N}} \ = \  P_N{}^R \phi_R
\,, ~~ \phi_{\bar N} \ = \ \bar P_N{}^R
\phi_R\,,
~~~\hbox{with} ~~  \phi_R \ = \ -2 \bigl( \partial_R d +  ( P\partial^Q  P)_{[QR]}  \bigr) \;.
\ee

Let us now resume the analysis of equations (\ref{fortraceeqns}).
A solution of these equations is of the form
\be
\label{anstrlkfk}
\begin{split}
\Gamma_{\nin{M}\nin{N}\nin{K}} \ &= \
 \alpha  P_{M[N} P_{K]}{}^Q  \phi_Q
 \ = \ \alpha \,
P_M{}^R
 P_{[N}{}^L P_{K]}{}^Q \, \eta_{RL}
  \phi_{{Q}}  \,,\\[1.0ex]
\Gamma_{\bar M\bar N\bar K} \ &= \
  \alpha  \bar P_{M[N} \bar P_{ K]}{}^Q\,  \phi_Q
 \ = \ \alpha \,
\bar P_M{}^R
\bar P_{[N}{}^L \bar P_{K]}{}^Q \, \eta_{RL} \phi_{ Q}  \,,
\end{split}
\ee
where $\alpha$ is a constant to be determined. The last
right-hand side on each line
 was written to make it manifest that the $\Gamma$'s have
the correct projections.  Note that this ansatz, as required,
satisfies constraints (1) and (2):
$\Gamma_{\nin{M}\nin{N}\nin{K}} = - \Gamma_{\nin{M}\nin{K}\nin{N}}$ and  $\Gamma_{\nin{M}\nin{N}\nin{K}} + \Gamma_{\nin{N}\nin{K}\nin{M}} + \Gamma_{\nin{K}\nin{M}\nin{N}}
= 0$.
The coefficient $\alpha$ is determined by contraction.  We get
\be
\eta^{MK} \Gamma_{\nin{M}\nin{N}\nin{K}} \ = \  {1\over 2}  \alpha (1-D) \phi_{\nin{N}}
 = \phi_{\nin{N}}  \quad \to \quad   \alpha = {2\over 1-D} \,.
\ee
Back in (\ref{anstrlkfk}) and using (\ref{singleff}) the full solution is therefore
\be
\label{thetwopieces}
\begin{split}
\Gamma_{\,\nin{M}\,\nin{N}\,\nin{K}} \ &= \  - {2\over (D-1)}  P_{M[N} P_{K]}{}^R\, \phi_R
+ \tilde \Gamma_{\nin{M}\nin{N}\nin{K}} \,, \\[1.0ex]
\Gamma_{\bar M\bar N\bar K} \ &= \  - {2\over (D-1)}
\bar P_{ M[ N} {\bar P}_{K]}{}^R\, \phi_R
+ \tilde \Gamma_{\bar M\bar N\bar K} \,,
\end{split}
\ee
where $\tilde \Gamma$ is undetermined and satisfies
\be
\label{trace-undetermined}
\begin{split}
\eta^{MK} \,\tilde \Gamma_{\nin{M}\nin{N}\nin{K}} \ &= \ 0 \,, \\[1.0ex]
\eta^{MK} \,\tilde \Gamma_{\bar M\bar N\bar K} \ &= \ 0 \,.
\end{split}
\ee

\subsubsection{The full Christoffel connection}

To write a complete expression for the Christoffel connection we
begin by adding the two contributions in (\ref{thetwopieces}) and
use (\ref{singleff}) to find
\be
\begin{split}
\Gamma_{\nin{M}\nin{N}\nin{K}} +
\Gamma_{\bar M\bar N\bar K} \ =&\,  \   {4\over (D-1)} \Bigl( P_{M[N} P_{K]}{}^R
+\bar P_{ M[ N} {\bar P}_{K]}{}^R \Bigr)\bigl( \partial_R d +  (\bar P\partial^Q \bar P)_{[QR]}  \bigr)\\
& + \tilde \Gamma_{\nin{M}\nin{N}\nin{K}} + \tilde \Gamma_{\bar M\bar N\bar K}\,.
\end{split}\ee
The full connection is then given by (\ref{explrepr99998}) which we write as
\be\label{explrcooonenf}
 \begin{split}
 \Gamma_{MNK}   \ &= \,
 (\Gamma_{\nin{M}\,\nin{N}\bar K} +\Gamma_{\bar M\,\nin{N}\bar K})
-   (\Gamma_{\nin{M}\,\nin{K}\bar N}   +\Gamma_{\bar M\,\nin{K}\bar N} )
    \\[1.0ex]
&~~
-\Gamma_{\nin{N}\,\nin{K}\bar M}
+ \Gamma_{\nin{K}\,\nin{N}\bar M }
- \Gamma_{\bar K\, \nin{M} \bar N}
+\Gamma_{\bar N\,\nin{M}\bar K}\,
\\[1.0ex]
&~~+   {4\over (D-1)} \Bigl( P_{M[N} P_{K]}{}^R
+\bar P_{ M[ N} {\bar P}_{K]}{}^R \Bigr)\bigl( \partial_R d +  ( P\partial^Q  P)_{[QR]}  \bigr) \\[1.0ex]
&~~ + \tilde\Gamma_{\nin{M}\nin{N}\nin{K}}  +\tilde\Gamma_{\bar M\bar N\bar K} \;. \\[1.0ex]
 \end{split}
 \ee
The first two lines on the above right-hand side can be evaluated
using equations (\ref{SomeComp}).  These equations imply, for example,
that
\be
\Gamma_{\nin{M}\nin{N}\bar K} + \Gamma_{\bar M \nin{N} \bar K} \ = \ - (\bar P \partial_M P)_{KN} \,.
\ee
With this one quickly verifies that the first line in
the  right-hand side of (\ref{explrcooonenf}) simplifies down to
$-2(P\partial_{M}P)_{[NK]}$.  A computation of the second line
then yields
the complete result.
We write it as
\be\label{Connectionsplit}
\Gamma_{MNK} \ = \ \widehat\Gamma_{MNK} + \Sigma_{MNK}\,,
\ee
where $\widehat\Gamma_{MNK}$ is the determined part of the
connection,
 \be\label{Finalgamma}
 \begin{split}
  \widehat\Gamma_{MNK} \ = \ &-2(P\partial_{M}P)_{[NK]}-2\big(\bar{P}_{[N}{}^{P}\bar{P}_{K]}{}^{Q}-P_{[N}{}^{P}P_{K]}{}^{Q}\big)\partial_{P}P_{QM}\\
  &+\frac{4}{D-1}\big(P_{M[N}P_{K]}{}^{Q}+\bar{P}_{M[N}\bar{P}_{K]}{}^{Q}\big)\big(\partial_{Q}d+(P\partial^{P}P)_{[PQ]}\big)
 \;,
 \end{split}
 \ee
and $\Sigma_{MNK}$ is the undetermined part of the
connection:
\be
\label{Fconnection}
\Sigma_{MNK} \ = \  \tilde\Gamma_{\,\nin{M}\,\nin{N}\,\nin{K}} \ +\ \tilde\Gamma_{\bar M \bar N \bar K}  \,.
\ee
The result (\ref{Finalgamma}) is equivalent to the ansatz given in eq.~(15) of \cite{Jeon:2011cn}.
The $\Sigma_{MNK}$
satisfy the traceless condition in (\ref{trace-undetermined}).  Given
the symmetry properties of the connection, the trace taken on any two indices of the $\tilde \Gamma$'s vanishes. This completes our calculation of the connection.
We finally give the number of undetermined connection components. Since $P$ and $\bar{P}$ are rank-$D$ projectors, any
projected $O(D,D)$ index represents $D$ independent components.
The two undetermined $\tilde{\Gamma}$ can thus be viewed as taking values in the $(2,1)$ traceless $GL(D)$ Young tableau.
The total number of undetermined components is then found to be $\tfrac{2}{3}D(D+2)(D-2)$, which is equal to the value in Siegel's frame-like
formalism, see the discussion after eq.~(2.40) in \cite{Hohm:2010xe}.

\medskip
\noindent
We can rewrite the above
$\widehat{\Gamma}$ directly in terms of
${\cal H}$ and $d$.  Using the definition of the projectors a quick
calculation shows that
\be
(P\partial_M P)_{PQ} \ = \  {1\over 4} \Bigl( - \partial_M {\cal H}_{PQ}  +
 {\cal H}_{PK} \partial_M {\cal H}^K{}_Q\Bigr) \,.
 \ee
The first term on the right-hand side is symmetric in $P$ and $Q$ while the second term is actually antisymmetric in $P$ and $Q$. We thus have
\be
(P\partial_M P)_{[PQ]} \ = \  {1\over 4}
 {\cal H}_{PK} \partial_M {\cal H}^K{}_Q\,.
\ee
As a result, we obtain
\be
(P\partial^P P)_{[PQ]} \ = \   {1\over 4} \,  {\cal H}_{PK} \partial^P
 {\cal H}^K{}_Q \ = \  {1\over 4} \,  {\cal H}^{PM} \partial_M {\cal H}_{PQ}\,.
\ee
We can quickly work out the other projectors:
\be
\bar P_{[N}{}^P  \bar P_{K]}{}^Q
 - P_{[N}{}^P  P_{K]}{}^Q \ = \ {1\over 2} \Bigl(\, \delta_{[N}{}^P  {\cal H}_{K]}{}^Q
 + {\cal H}_{[N}{}^P  \delta_{K]}{}^Q\Bigr)\;,
\ee
\be
\bar P_{M[N}  \bar P_{K]}{}^Q
 +  P_{M[N}   P_{K]}{}^Q \ = \ {1\over 2} \Bigl( \eta_{M[N} \delta_{K]}{}^Q
 +  {\cal H}_{M[N}   {\cal H}_{K]}{}^Q  \Bigr)\;.
\ee
Back in the connection (\ref{Finalgamma})  we get
 \be\label{Finalgamma99}
 \begin{split}
  \widehat\Gamma_{MNK} \ = \ &\, {1\over 2} \, {\cal H}_{KQ} \partial_M {\cal H}^Q{}_N
  +{1\over 2}  \Bigl(\, \delta_{[N}{}^P  {\cal H}_{K]}{}^Q
 + {\cal H}_{[N}{}^P  \delta_{K]}{}^Q\Bigr)\partial_{P}{\cal H}_{QM}\\
  &+\frac{2}{D-1}
 \Bigl( \eta_{M[N} \delta_{K]}{}^Q
 +  {\cal H}_{M[N}   {\cal H}_{K]}{}^Q  \Bigr)
  \Bigl(\partial_{Q}d+ {1\over 4}
 {\cal H}^{PM} \partial_M {\cal H}_{PQ}\Bigr)
 \;.
 \end{split}
 \ee

\section{Analysis of the generalized Riemann tensor}

In this section we examine the components of the generalized
tensor ${\cal R}_{MNPQ}$ using the projected barred and
under-barred indices.   We show that the projections in which
undetermined connections drop out vanish identically.
There are four non-vanishing
projections, as detailed
in equation~(\ref{freaklist2}).  We then show how the Ricci
and scalar generalized curvatures arise from ${\cal R}_{MNPQ}$
by taking contractions that make all undetermined connections
disappear.  An analysis of the invariant action allows us to show
that there is a single generalized Ricci curvature and to prove
differential Bianchi identities.

\subsection{The components  of the Riemann tensor}\label{thecompon}

Before we begin the detailed discussion of the various components
of the Riemann tensor, we examine a useful property that follows
from the covariant constancy of the projectors.
This property implies that:
 \be
  [\nabla_{M},\nabla_{N}]P_{K}{}^{L}V_{L} \ = \ P_{K}{}^{L} [\nabla_{M},\nabla_{N}]V_{L}\;,
 \ee
so that expanding the commutators according to (\ref{commutator}) we get
 \be
  -R_{MNK}{}^{P}P_{P}{}^{L}V_{L}-T_{MN}{}^{P}\nabla_{P}(P_{K}{}^{L}V_{L}) \ = \
  -P_{K}{}^{L}R_{MNL}{}^{P}V_{P}-P_{K}{}^{L}T_{MN}{}^{P}\nabla_{P}V_{L}\;.
 \ee
Using the covariant constancy again we see that the torsion terms cancel on both sides.
Relabeling indices and dropping the $V$'s we obtain
 \be
  R_{MNKP}P^{P}{}_{L} \ = \ R_{MNPL}P^{P}{}_{K}\;.
 \ee
Multiplying by $\bar{P}^{K}{}_{Q}$ we see that the above right-hand side vanishes due to $P\bar{P} = 0$.  We therefore find that
 \be\label{mixednull}
  R_{MNKP}\bar{P}^{K}{}_{Q} P^{P}{}_{L}    \ =  \ 0\quad \to \quad
  R_{MN\bar Q \,\nin{L}} \ =  \ 0\,. \ee
A curvature $R$ with mixed projections on the last two indices
vanishes.

\medskip
In order to find out which components of the curvature depend on undetermined connections we use
 the variation formula (\ref{varcalR}) and the split
 (\ref{Connectionsplit})
of the connection into a determined piece $\widehat \Gamma$ and an undetermined piece $\Sigma$.  We find
 \be\label{covRexp}
  \begin{split}
   {\cal R}_{MNKL} \ = \ &\ \widehat{\cal R}_{MNKL}  \ +\ 2\, \widehat{\nabla}_{[M}\Sigma_{N]KL}\ + \ 2\, \widehat{\nabla}_{[K}\Sigma_{L]MN} \\[1.0ex]
   &\hskip-7pt +\ 2\,\Sigma_{[M|QL|}\Sigma_{N]K}{}^{Q}\ +\ 2\,\Sigma_{[K|QN|}\Sigma_{L]M}{}^{Q}\ +\ \Sigma_{QMN}\Sigma^{Q}{}_{KL}\;.
  \end{split}
 \ee
In here all hatted quantities are ones that use
$\widehat \Gamma$.

\medskip
Let us now consider possible components of the
projected curvatures ${\cal R}$.
There is one ${\cal R}$ with all indices under-barred and one ${\cal R}$ with
all indices barred  -- a type $(4,0)$ curvature in the notation
introduced in (\ref{type-defined}).
 With three under-barred indices and one barred one
there is just one ${\cal R}$ since the barred index can always
be chosen to be the last by using the pair exchange symmetry and
the antisymmetry in the last two indices.
The same is true for the ${\cal R}$
with three barred indices and one under-barred one.  Finally for an
${\cal R}$ with two indices of each type there are two configurations:
one in which  the first and last two indices are of the same type,
and one where they are not.  In summary,
 \be\label{freaklist}
  {\cal R}_{\nin{M}\nin{N}\nin{K}\nin{L}}\;,\quad
  {\cal R}_{\nin{M}\nin{N}\nin{K}\bar{L}}\;, \quad
 {\cal R}_{\nin{M}\bar{N}\,\nin{K}\bar{L}}\;, \quad
   {\cal R}_{\nin{M}\nin{N}\bar{K}\bar{L}}\;, \quad
 {\cal R}_{\nin{M}\bar{N}\bar{K}\bar{L}}\;, \quad
 {\cal R}_{\bar{M}\bar{N}\bar{K}\bar{L}}\;.
 \ee
The two type $(2,2)$ curvatures
are not independent. The algebraic
Bianchi identity (\ref{algebraicBianchi})  gives
 \be
  0 \ = \ {\cal R}_{\nin{M}\nin{N}\bar{K}\bar{L}} + {\cal R}_{\nin{N}\bar{K}\,\nin{M}\bar{L}} + {\cal R}_{\bar{K}\nin{M}\nin{N}\bar{L}}
 ~~\to ~~{\cal R}_{\nin{M}\nin{N}\bar{K}\bar{L}}  \ = \ {\cal R}_{\nin{M}\bar{K}\,\nin{N}\bar{L}} -{\cal R}_{\nin{N}\bar{K}\,\nin{M}\bar{L}}\,,
 \ee
showing that  the third curvature in (\ref{freaklist}) determines the fourth.
The third structure, using  definition (\ref{covRiem}), is given by
\be
\label{Thirdzero}
   {\cal R}_{\,\nin{M}\bar{N}\,\nin{K}\bar{L}} \ = \ R_{\,\nin{M}\bar{N}\,\nin{K}\bar{L}}+R_{\,\nin{K}\bar{L}\, \nin{M}\bar{N}}+\Gamma_{Q\,\nin{M}\bar{N}}
   \Gamma^{Q}{}_{\nin{K}\bar{L}}\,.
 \ee
The first two terms vanish because of  (\ref{mixednull}) and the
last one contains pieces of the connection determined in
(\ref{someprojection}):
 \be\label{Thirdzero}
   {\cal R}_{\nin{M}\bar{N}\,\nin{K}\bar{L}} \ = \ \Gamma_{Q\,\nin{M}\bar{N}}\Gamma^{Q}{}_{\nin{K}\bar{L}}
   \ = \ (\bar{P}\partial_{Q}P)_{MN}\,(\bar{P}\partial^{Q}P)_{KL} \ = \ 0\;,
 \ee
using the strong constraint. The vanishing of this third structure then
implies the vanishing of the fourth, as remarked above:
\be
{\cal R}_{\,\nin{M}\nin{N}\bar K \bar L} \ = \ 0 \,.
\ee

With (\ref{covRexp}) it is now easy to see that the first two and last two in (\ref{freaklist}) depend on the
undetermined connections.
In fact, for
${\cal R}_{\,\nin{M}\nin{N}\nin{K}\nin{L}}$ we use  (\ref{covRexp})
together with (\ref{Fconnection}) to get
 \be\label{covRexpprime}
  \begin{split}
   {\cal R}_{\nin{M}\nin{N}\nin{K}\nin{L}} \ = \ &\ \widehat{\cal R}_{\nin{M}\nin{N}\nin{K}\nin{L}}  \ +\ 2\, \widehat{\nabla}_{[\,\nin{M}}\,\tilde\Gamma_{\nin{N}]\,\nin{K}\nin{L}}\ + \ 2\, \widehat{\nabla}_{[\,\nin{K}}\tilde\Gamma_{\nin{L}\,]\,\nin{M}\nin{N}} \\[1.0ex]
   &\hskip-7pt +\ 2\,\tilde\Gamma_{[\,\nin{M}|\,\nin{Q}\,\nin{L}|}\tilde\Gamma_{\,\nin{N}\,]\,\nin{K}}{}^{\nin{Q}}\
   +\ 2\,\tilde\Gamma_{[\,\nin{K}|\,\nin{Q}\,\nin{N}\,|}
   \tilde\Gamma_{\nin{L}\,]\,\nin{M}}{}^{\nin{Q}}\
   +\ \tilde\Gamma_{\nin{Q}\,\nin{M}\nin{N}}\,\tilde\Gamma^{\,\nin{Q}}{}_{\,\nin{K}\nin{L}}\;.
  \end{split}
 \ee
In here, projected indices on covariant derivatives
are defined as usual:  $\widehat{\nabla}_{\bar L} \equiv
\bar{P}_{L}{}^{Q}\widehat{\nabla}_{Q}$.
We note that all $\Sigma_{MNK}$ in (\ref{covRexp})
were replaced by $\tilde{\Gamma}_{\nin{M}\nin{N}\nin{K}}$
because the projectors discard the $\tilde{\Gamma}_{\bar{M}\bar{N}\bar{K}}$ components.  Note that the summed index $Q$ only receives
contributions from the under-barred values.  Analogous remarks apply for
the fully barred structure ${\cal R}_{\bar M\bar N\bar K\bar L}$.

For the second curvature in the list,
the type $(3,1)$ tensor ${\cal R}_{\nin{M}\nin{N}\nin{K}\bar{L}}$,
all $\Sigma^2$ terms vanish because in each of
them one $\Sigma$ has mixed barred/under-barred projections and there
are no such undetermined connections. From the $\nabla\Sigma$ type terms, one  survives:
 \be\label{31type}
 {\cal R}_{\nin{M}\nin{N}\nin{K}\bar{L}} \ = \ \widehat{\cal R}_{\nin{M}\nin{N}\nin{K}\bar{L}}
  -\widehat{\nabla}_{\bar L}\tilde\Gamma_{\,\nin{K}\nin{M}\nin{N}}\,.
 \ee
We thus see that ${\cal R}_{\nin{M}\nin{N}\nin{K}\bar{L}}$
involves undetermined connections.
Similarly, we find for the $(1,3)$ type structure
\be
\label{13type}
{\cal R}_{\,\nin{M} \bar N \bar K \bar L} \ = \ \widehat
{\cal R}_{\nin{M} \bar N \bar K \bar L}  \ + \ \widehat\nabla_{\,\nin{M}}
 \tilde\Gamma_{\bar N \bar K \bar L} \,.
\ee

Our analysis shows that the list (\ref{freaklist}) has become
 \be\label{freaklist2}
 \begin{split}
 & {\cal R}_{\nin{M}\nin{N}\nin{K}\,\nin{L}}~~  \hbox{~~contains undetermined connections,}\\
& {\cal R}_{\nin{M}\nin{N}\nin{K}\bar{L}}~~  \hbox{~~contains undetermined connections,}\\
& {\cal R}_{\nin{M}\bar{N}\,\nin{K}\bar{L}} \ = \ 0 \;, \\
 &  {\cal R}_{\nin{M}\nin{N}\bar{K}\bar{L}} \ \, =  \ 0 \;,\\
 &{\cal R}_{\nin{M}\bar{N}\bar{K}\bar{L}}~~\hbox{~~contains undetermined connections,} \\
 &{\cal R}_{\bar{M}\bar{N}\bar{K}\bar{L}}~~\hbox{~~contains undetermined connections.}
\end{split}
 \ee
Thus, there is no Riemann tensor
in terms of the physical fields.

\subsection{Generalized Ricci and scalar curvatures}

Undetermined connection components
can drop out from traces of curvatures. In fact,  we can define a scalar
curvature  and a Ricci tensor.   A naive candidate for
the scalar curvature is $ {\cal R}_{MN}{}^{MN}$.  Expanding
the contractions in projected indices we have,
\be
 {\cal R}_{MN}{}^{MN} \ = \  \eta^{MK} \eta^{NL} \,{\cal R}_{MNKL}
  \ = \ {\cal R}_{\,\nin{M}\nin{N}}{}^{\,\nin{M}\nin{N}} + {\cal R}_{\bar{M}\bar{N}}{}^{\bar{M}\bar{N}}   +2 \, {\cal R}_{\,\nin{M}\bar{N}}{}^{\,\nin{M}\bar{N}}\,.
\ee
The last term on the right-hand side vanishes
by (\ref{Thirdzero}),
so that we have
\be
\label{fulltrace=2}
 {\cal R}_{MN}{}^{MN} \ = \   {\cal R}_{\,\nin{M}\nin{N}}{}^{\,\nin{M}\nin{N}} + {\cal R}_{\bar{M}\bar{N}}{}^{\bar{M}\bar{N}}  \ \,.
\ee
Recall from (\ref{contract-project})
that contractions on projected indices are implemented
by contractions against the appropriate projector, so that
\be
\label{contract-proje}
\begin{split}
 {\cal R}_{\,\nin{M}\nin{N}}{}^{\,\nin{M}\nin{N}}  \ = \ P^{MK}P^{NL}{\cal R}_{MNKL} \,,
\qquad  {\cal R}_{\bar{M}\bar{N}}{}^{\bar{M}\bar{N}}  \ &= \ \bar P^{MK}\bar P^{NL}{\cal R}_{MNKL} \,.
 \end{split}
\ee
Back to (\ref{fulltrace=2}) we compute ${\cal R}_{MN}{}^{MN}$
 directly from the definition
(\ref{covRiem}) and from (\ref{conRieTor}):
 \be\label{zerotrace}
 \begin{split}
  {\cal R}_{MN}{}^{MN} \ &= \ 2R_{MN}{}^{MN}+\Gamma_{MNK}\Gamma^{MNK} \\
  \ &= \ 4\partial_{M}\Gamma^{M}+ 2\,
  \Gamma_{M}\Gamma^{M}+2\,\Gamma_{MNK} \Gamma^{KMN}+ \Gamma_{MNK}
 \Gamma^{MNK} \ \equiv \ 0 \;.
 \end{split}
 \ee
The first two terms on the right-hand side vanish using
$\Gamma_{M}\sim \partial_{M}d$ and the strong constraint.
The rest of the terms on the right-hand side vanish too:
 \be\label{zerotrace99}
 \begin{split}
  {\cal R}_{MN}{}^{MN} \ &= \
 \,\Gamma_{MNK}\Bigl(  \Gamma^{KMN}+\Gamma^{KMN} +
 \Gamma^{MNK}\Bigr) \\[1.0ex]
 &= \
 \,\Gamma_{MNK}\Bigl(  \Gamma^{KMN}-\Gamma^{NMK} +
 \Gamma^{MNK}\Bigr)\ = \ 0 \;,
 \end{split}
 \ee
because of $\Gamma_{[MNK]}=0$. The vanishing of $ {\cal R}_{MN}{}^{MN}$ is consistent with the vanishing of the flat-index combination
${\cal R}_{AB}{}^{AB}$ in Siegel's formalism \cite{Siegel:1993th}.
Equation (\ref{fulltrace=2}) and the vanishing of $ {\cal R}_{MN}{}^{MN}$ suggest that
we have to contract the fully projected tensors.  We thus define the
scalar curvature ${\cal R}$ by
 \be\label{Ricciscalar}
  {\cal R} \ \equiv \ {\cal R}^{\,\nin{M}\nin{N}}{}_{\nin{M}\nin{N}} \ = \ -{\cal R}^{\bar{M}\bar{N}}{}_{\bar{M}\bar{N}}\;.
 \ee
We now use (\ref{covRexpprime}) to show that the undetermined
connections drop out of ${\cal R}$. Let us do one contraction first.
The tracelessness of the $\tilde\Gamma$ (on any pair of indices)  implies that
\be
\label{symcontraction}
\begin{split}
\eta^{NL} {\cal R}_{\,\nin{M}\,\nin{N}\,\nin{K}\,\nin{L}}
 \ =& \  \widehat{\cal R}_{\,\nin{M}\,\nin{N}\,\nin{K}}{}^{\,\nin{N}}
\, -\, \widehat\nabla^{\nin{L}} \bigl(  \tilde\Gamma_{\,\nin{M}\,\nin{K}\,\nin{L}}
 + \tilde\Gamma_{\,\nin{K}\,\nin{M}\,\nin{L}} \bigr) \\[1.0ex]
 & + \tilde\Gamma_{\,\nin{M} \,\nin{Q}} {}^{\,\nin{L}}  \, \,
 \tilde\Gamma_{\,\nin{L} \, \nin{K} } {}^{\nin{Q}}
 \ + \ \tilde\Gamma_{\,\nin{K} \,\nin{Q}} {}^{\,\nin{L}}  \, \,
 \tilde\Gamma_{\,\nin{L} \, \nin{M} } {}^{\nin{Q}}
 \ + \  \tilde\Gamma_{\,\nin{Q} \,\nin{M}\,\nin{L}}  \, \,
 \tilde\Gamma{}^{\,\nin{Q}}{}_{\nin{K}} {}^{\nin{L}} \,.
\end{split}
\ee
A few undetermined connection coefficients dropped out but several
remain.  After the second contraction with $\eta^{MK}$ we get
only $\tilde \Gamma \tilde\Gamma$ terms
that survive, but they
add up to zero:
 \be
 \label{dropoutfromRunder}
 \begin{split}
   {\cal R} \ &= \  \widehat{\cal R}^{\,\nin{M}\nin{N}}{}_{\nin{M}\nin{N}}
   +\tilde\Gamma_{\nin{M}\,\nin{Q}\,\nin{L}}
   \,\tilde\Gamma^{\,\nin{L}\,\nin{M}\,\nin{Q}}
    +\tilde\Gamma_{\,\nin{K}\,\nin{Q}\,\nin{N}}\,
    \tilde\Gamma^{\,\nin{N}\nin{K}\,\nin{Q}}
   +\tilde\Gamma_{\,\nin{Q}\,\nin{M}\,\nin{N}}\,
   \tilde\Gamma^{\,\nin{Q}\,\nin{M}\,\nin{N}} \\[0.5ex]
   \ &= \ \widehat{\cal R}^{\,\nin{M}\nin{N}}{}_{\nin{M}\nin{N}}
   +\tilde\Gamma_{\nin{M}\,\nin{Q}\, \nin{L}}
  \, \bigl(\,\tilde\Gamma^{\,\nin{L}\,\nin{M}\,\nin{Q}}
   +\tilde\Gamma^{\,\nin{M}\,\nin{Q}\,\nin{L}}
   +\tilde\Gamma^{\,\nin{L}\,\nin{M}\,\nin{Q}}\,\bigr) \\[0.5ex]
   \ &= \  \widehat{\cal R}^{\,\nin{M}\nin{N}}{}_{\nin{M}\nin{N}}\;,
 \end{split}
 \ee
using the generalized torsion constraint.
The undetermined connections dropped
 out and there
is a well-defined scalar curvature ${\cal R}$.
It must be proportional to the scalar curvature defined in \cite{Hohm:2010pp}.
One may fix the normalization by inserting the explicit connection components, say, focusing on the
dilaton-dependent terms. We then find that (\ref{Ricciscalar}) equals the curvature scalar
defined in eq.~(4.24) in  \cite{Hohm:2010pp}.

Equation (\ref{symcontraction}) shows that we cannot get
a well-defined Ricci tensor with two under-barred (or two barred) indices.
The Ricci tensor is of type $(1,1)$,
and we can define such an object by contraction with $\eta$ of a curvature with $(1,3)$
or $(3,1)$ index structure.  We define the following objects starting
with the
$(3,1)$ index structure:
\be
\label{introricci}
\begin{split}
  {\cal R}_{\nin{M}\bar{N}} \ \equiv  &\,\ {\cal R}_{\nin{K}\nin{M}\bar{N}}{}^{\,\nin{K}} \ = \ \eta^{KL}\,
 {\cal R}_{\nin{K}\,\nin{M}\bar{N}\,\nin{L}}\;,\\[1.0ex]
  {\cal R}_{\bar{N}\,\nin{M}} \ \equiv  &\,\ {\cal R}_{\nin{K}\bar{N}\,\nin{M}}{}^{\,\nin{K}} \ = \ \eta^{KL}\,
 {\cal R}_{\nin{K}\bar{N}\,\nin{M}\,\nin{L}}\;.
 \end{split}
\ee
In fact, the Bianchi identity implies they are equal:
\be\label{EqualRic}
 {\cal R}_{\bar{N}\,\nin{M}} \ = \ {\cal R}_{\nin{K}\bar{N}\,\nin{M}}{}^{\,\nin{K}}
\ = \ - {\cal R}_{\bar{N}\,\nin{M}\nin{K}}{}^{\,\nin{K}}-
{\cal R}_{\,\nin{M}\nin{K}\bar{N}}{}^{\,\nin{K}} \ = \
{\cal R}_{\nin{K}\,\nin{M}\bar{N}}{}^{\,\nin{K}}\ = \  {\cal R}_{\nin{M}\bar{N}} \,.\ee
This is the ``symmetry" property of the Ricci curvature.  Most importantly,
undetermined connections do not appear in the Ricci curvature.
Indeed, starting from the definition (\ref{introricci})
and using (\ref{31type})
we have
\be
 {\cal R}_{\nin{M}\bar{N}}\ = \  \eta^{KL}\,
 {\cal R}_{\,\nin{M}\nin{K}\,\nin{L}\,\bar{N}} \ = \
  \eta^{KL}\,
\bigl( \widehat
{\cal R}_{\nin{M}  \nin{K}  \,\nin{L}\bar N}  \ - \
\widehat\nabla_{\bar{N}}
\, \tilde\Gamma_{ \nin{L}\, \nin{M}\nin{K}}\bigr) \ = \   \eta^{KL}\,
 \widehat
{\cal R}_{\nin{M}  \nin{K} \nin{L}\bar N}\,.
\ee
We will show in the following subsection that the Ricci tensor
defined by contraction of the $(1,3)$ index structure is identical to the one
obtained here.

\subsection{Invariant action and differential Bianchi identities}
After having defined a generalized curvature scalar ${\cal R}$ we can define an invariant
action for double field theory. It reads
 \be\label{actionprin99}
  S \ = \ \int dxd\tilde{x}\,e^{-2d}\,{\cal R} \ = \  \int dxd\tilde{x}\,e^{-2d}\,
  {\cal R}_{\nin{M}\nin{N}}{}^{\,\nin{M}\nin{N}}\ = \
  \int dx d\tilde{x} e^{-2d} P^{MK}P^{NL}{\cal R}_{MNKL} \;,
 \ee
 where we recalled (\ref{contract-proje}). Since the undetermined
 pieces of the connection drop out (see (\ref{dropoutfromRunder})), we have
  \be\label{actionprin}
  S \ =  \
  \int dx d\tilde{x}\, e^{-2d} P^{MK}P^{NL}\widehat{\cal R}_{MNKL} \;.
 \ee
Of course, on account
 of (\ref{Ricciscalar}) we also have
\be\label{actionprin22}
  S \ = \  -\int dxd\tilde{x}\,e^{-2d}\,
  {\cal R}_{\bar{M}\bar{N}}{}^{\bar{M}\bar{N}}\ = \
 - \int dx d\tilde{x} e^{-2d} \bar P^{MK}\bar P^{NL}\widehat {\cal R}_{MNKL} \;.
 \ee
It turns out that further Bianchi-type identities for the Ricci tensor and curvature scalar can be conveniently derived
using the invariance properties of this action.

We start by discussing the variational principle based on (\ref{actionprin}).
For earlier discussions of the general variation in double field theory see \cite{Hohm:2010jy,Hohm:2010pp,Kwak:2010ew,Hohm:2010xe,Jeon:2011cn}.  
Variations of the generalized metric imply variations of
 $P$ or $\bar{P}$.
 In fact we can think of $P$ and $\bar P$ as the field variables to
 be varied since the connection is written in terms of these
 projectors (see (\ref{Finalgamma})).  We must then
 take into account that these are
constrained to satisfy $P^2=P$, $\bar{P}^2=\bar{P}$ and $P\bar{P}=0$. Thus if we shift
$P^{\prime}=P+\delta P$ we need to satisfy
 \be
  (P^{\prime})^2 \ = \ P+P\delta P+\delta P P \ \equiv \ P^{\prime} \ = \ P+\delta P\;,
 \ee
and similarly for $\bar{P}$. Thus, we have the constraint
 \be
  \delta P \ = \   P\delta P+\delta P P\;,
 \ee
and similarly for $\bar{P}$.   Acting on both sides with $P$ from the left and the right we quickly see that $P\delta P P = 0$.
Moreover, we also see that
$ \bar{P}\delta P\bar{P} = 0$.
Finally, when acting from the left with $P$ and the right with $\bar{P}$, or vice versa,
we get trivially satisfied identities that  imply that $P\delta P\bar{P}$ and $\bar{P}\delta P P$ are unconstrained.
Thus, we can write the variation in terms of two unconstrained matrices ${\cal M}$
and ${\cal N}$ as follows
 \be\label{delP0}
  \delta P \ = \ \bar{P}{\cal M}P+P{\cal N}\bar{P} \ = \ -\delta \bar{P}\;,
 \ee
where the last condition follows from $P + \bar P = 1$.
Since $P$ and $\bar{P}$ are symmetric, $\delta P$ and $\delta\bar{P}$ should be symmetric too,
requiring that
${\cal M}^T={\cal N}$. Thus, the most general variations of $P$ and
$\bar P$ consistent with the constraints are
 \be\label{delP}
  \delta P \ = \ \bar{P}{\cal M}P+P{\cal M}^{T}\bar{P} \ = \ -\delta \bar{P}\;.
 \ee

Let us now consider the general variation of the action (\ref{actionprin})
for variations $\delta P$ and $\delta d$.  Of course such variations result
in variations $\delta\, \widehat \Gamma$ of the determined parts of the
connection.  The undetermined parts need not be varied since they and
their variations drop out of the action.  We thus get
 \be\label{GenrVarrr}
 \begin{split}
   \ \delta S \ &= \  \delta  \int dxd\tilde{x}\,e^{-2d}\,P^{MK}P^{NL}
  \widehat {\cal R}_{MNKL} \\
   \ &= \
   \int dxd\tilde{x}\,e^{-2d}\left(-2\delta d\,{\cal R}+2\,\delta P^{MK}P^{NL} \widehat{\cal R}_{MNKL}
   +4P^{MK}P^{NL}\widehat{\nabla}_{[M}\delta\widehat\Gamma_{N]KL}\right)\;,
 \end{split}
 \ee
where we employed (\ref{varcalR})
since  this relation holds for any shift of the
connection.
The covariant derivative in
 $\widehat \nabla\delta \widehat\Gamma$ can be partially integrated:  it ignores the dilaton density and gives zero
acting on the $P$'s (note that both $\nabla$ and $\widehat \nabla$ have such
properties).
This term is therefore a total derivative, in complete analogy
to standard Einstein gravity. The variation of $d$ then implies the vanishing of the scalar curvature, ${\cal R}=0$.
This is a well-known result in double field theory~\cite{Hohm:2010jy,Hohm:2010pp}, but here we
understand more clearly
 why the variations of $d$ inside ${\cal R}$ add up to a total derivative.

We focus on the remaining variation which reads with (\ref{delP})
 \be\label{varunbarred}
 \begin{split}
  \delta S
   \ &= \ 2\int dxd\tilde{x}\,e^{-2d}\left(\bar{P}^{MP}{\cal M}_{PQ}P^{QK}+P^{MQ}{\cal M}_{PQ}\bar{P}^{PK}\right)
  P^{NL}\widehat{\cal R}_{MNKL} \\
  \ &= \ 2\int dxd\tilde{x}\,e^{-2d}{\cal M}_{PQ}
  \left(\widehat{\cal R}^{\bar{P}\nin{L}\,\nin{Q}}{}_{\nin{L}}
  +\widehat{\cal R}^{\,\nin{Q}\,\nin{L}\,\bar{P}}{}_{\nin{L}}\right)
   \\
   \ & = \ - 2\int dxd\tilde{x}\,e^{-2d}{\cal M}_{PQ}
  \left({\cal R}^{\,\nin{L}\bar{P}\,\nin{Q}}{}_{\nin{L}}
  +{\cal R}^{\,\nin{L}\,\nin{Q}\bar{P}}{}_{\nin{L}}\right)
  \ = \
  -4\int dxd\tilde{x}\,e^{-2d}{\cal M}^{NM}\,{\cal R}^{\,\nin{K}}{}_{\nin{M}\bar{N}\nin{K}} \;,
 \end{split}
 \ee
where we were able to remove
 the hats at the point where we know
all undetermined connections drop out.
In the last step we used
(\ref{EqualRic}) and relabeled indices.
Thus, we get the field equation
 \be
  {\cal R}_{\,\nin{M}\bar{N}} \ \equiv \ {\cal R}^{\,\nin{K}}{}_{\nin{M}\bar{N}\,\nin{K}} \ = \ 0\;,
 \ee
recovering the Ricci tensor defined above.

 An alternative definition for the Ricci tensor is obtained by tracing
the curvature with
$(1,3)$ index structure
(one under-barred, three barred).
We will show now that the resulting object,
$\,{\cal R}_{\bar{K}\nin{M}\bar{N}}{}^{\,\bar{K}}$,
does not provide a new tensor.
To this end we vary the alternative form of the action
indicated in (\ref{actionprin22}):
  \be
      \delta S \ = \  -\delta  \int dxd\tilde{x}\,e^{-2d}\,\bar{P}^{MK}\bar{P}^{NL}\widehat{\cal R}_{MNKL}\;.
  \ee
Using $\delta \bar{P}=-\delta P$ we
arrive at
 \be
 \begin{split}
  \delta S \ &= \ 2\int dxd\tilde{x}\,e^{-2d}\,\delta P^{MK}\bar{P}^{NL}
  \widehat{\cal R}_{MNKL} \\
   \ &= \ - 2\int dxd\tilde{x}\,e^{-2d}\, \left(
  \bar{P}^{MP}{\cal M}_{PQ}P^{QK}
  +P^{MQ}{\cal M}_{PQ}\bar{P}^{PK}\right)
  \widehat {\cal R}^{\,\bar L}{}_{MK\bar{L}} \\
    \ &= \ - 2\int dxd\tilde{x}\,e^{-2d}\,  {\cal M}_{PQ} \left(
  {\cal R}^{\bar L \bar P\, \nin{Q}} {}_{\bar{L}}  +
   {\cal R}^{\bar L \, \nin{Q}\,\bar P} {}_{\bar{L}}  \right)
  \\
 \ &= \ -
 4\int dxd\tilde{x}\,e^{-2d}{\cal M}_{PQ}\,
 {\cal R}^{\bar L \, \nin{Q}\,\bar P} {}_{\bar{L}}
   \ = \ -4\int dxd\tilde{x}\,e^{-2d}{\cal M}^{NM}{\cal R}^{\bar{K}}{}_{\nin{M}\bar{N}\bar{K}}
   \;.
 \end{split}
 \ee
Here we combined the two terms in the third
line  using  the analogue of (\ref{EqualRic})
and removed the hats, since the objects in question are
well-defined.
As this variation must agree with the
variation (\ref{varunbarred}) for all ${\cal M}$ we
conclude
 \be
  {\cal R}^{\,\nin{K}}{}_{\nin{M}\bar{N}\,\nin{K}} \ =\  {\cal R}^{\bar{K}}{}_{\nin{M}\bar{N}\bar{K}} \;,
 \ee
proving that there is a single generalized Ricci tensor.

Let us relate the above definition of a Ricci tensor to a similar tensor defined  in \cite{Hohm:2010pp},
where we considered the variation of ${\cal H}$ rather than $P$.
The variation (\ref{delP}) implies the following variation for ${\cal H}$
 \be\label{DelH}
  \delta {\cal H} \ = \ -2\big(\bar{P}{\cal M}P+P{\cal M}^{T}\bar{P}\big)\;,
 \ee
where we used (\ref{projIntro}) and $\delta\eta=0$.  Up to the
factor of $-2$ this coincides with the
variation given in eq.~(4.54) in \cite{Hohm:2010pp} if we assume ${\cal M}$ to be symmetric.
The variation considered in \cite{Hohm:2010pp} was not the most general,
because ${\cal M}$ need not be symmetric, but it can be proved that the resulting field equations are equivalent to
the ones obtained for general ${\cal M}$. To see this, consider a general action $S$ based on a Lagrangian
${\cal L}(P)$ that we view as a function of $P$ (suppressing the dependence on other fields). Using (\ref{delP}),  its variation
with respect to $P$ then reads
 \be
 \begin{split}
  \delta S \ &= \
  \int dxd\tilde{x}\,e^{-2d} \,\frac{\delta {\cal L}}{\delta P_{KL}}
  \big(\bar{P}_{KM}{\cal M}^{MN}P_{NL}+P_{KN}{\cal M}^{MN}\bar{P}_{ML}\big) \\
  \ &= \ \int dxd\tilde{x}\, e^{-2d} \,{\cal M}^{MN}\left(\bar{P}_{MK}P_{NL}+\bar{P}_{ML} P_{NK}\right)\frac{\delta {\cal L}}{\delta P_{KL}} \\
  \ &= \ 2\int dxd\tilde{x}\, e^{-2d} \,{\cal M}^{MN}\,\bar{P}_{MK}P_{NL} \frac{\delta {\cal L}}{\delta P_{KL}}\;,
 \end{split}
 \ee
where we used in the last step the symmetry of $\delta {\cal L}/\delta P_{KL}$. As ${\cal M}_{MN}$ is unconstrained,
the field equations read
 \be
  E_{MN} \ \equiv \ \bar{P}_{MK}P_{NL} \frac{\delta {\cal L}}{\delta P_{KL}} \ = \ 0\;.
 \ee
An interesting property of tensors defined like this is that they vanish if and only if their symmetric
projection $E_{(MN)}$ vanishes. For suppose
 \be
  E_{(MN)} \ = \ \frac{1}{2}\left(\bar{P}_{MK}P_{NL} +\bar{P}_{NK}P_{ML}\right) \frac{\delta {\cal L}}{\delta P_{KL}} \ = \ 0\;.
 \ee
We can then contract with $\bar{P}_{R}{}^{M}$, after which the second term vanishes by $P\bar{P}=0$,
implying $E_{RN}=0$,
as we wanted to show. Thus, the field equations obtained by variation with a symmetric or general ${\cal M}$
are equivalent.

After this preliminary discussion it is straightforward to relate the Ricci tensor in \cite{Hohm:2010pp}
to the one discussed here. We consider the variation of the action (\ref{actionprin99}) under (\ref{DelH}) (or, equivalently, (\ref{delP0})),
with ${\cal M}$ symmetric,
 \be\label{VarStep}
  \delta S \ \equiv \ -2\int dxd\tilde{x}e^{-2d}\,{\cal M}^{MN}{\cal R}_{MN} \ = \
  -4\int dxd\tilde{x}e^{-2d}\,{\cal M}^{MN}P_{M}{}^{K}\bar{P}_{N}{}^{L}{\cal R}_{\bar{P}KL}{}^{\bar{P}}\;.
 \ee
The first equality
can be seen as the definition of ${\cal R}_{MN}$, where we
included a
factor of $-2$ such that the variation and hence the tensor
${\cal R}_{MN}$ have the same normalization as in \cite{Hohm:2010pp}.
For the second equality
we used (\ref{varunbarred}).
Since we assumed ${\cal M}$ to be symmetric, ${\cal R}_{MN}$ is symmetric, too, and from (\ref{VarStep}) given by
 \be\label{genRicci99}
  {\cal R}_{MN} \ = \ \big(P_{M}{}^{K}\bar{P}_{N}{}^{L}+P_{N}{}^{K}\bar{P}_{M}{}^{L}\big){\cal R}_{\bar{P}KL}{}^{\bar{P}}\,.
 \ee
Writing the right-hand side out in terms of projected indices
and using (\ref{EqualRic}) we get
 \be\label{genRicci}
  {\cal R}_{MN} \ = \
  {\cal R}_{\,\nin{M}\bar{N}}+{\cal R}_{\bar{M}\,\nin{N}}\,.
 \ee
The generalized tensor ${\cal R}_{MN}$ thus obtained
has no projected indices.
We can think of ${\cal R}_{\,\nin{M}\bar{N}}$ and ${\cal R}_{\bar{M}\,\nin{N}}$  as the projections of ${\cal R}_{MN}$.
The symmetric field equation ${\cal R}_{MN}=0$
is equivalent to ${\cal R}_{\,\nin{M}\bar{N}}=0$.

\medskip

We close this section by deriving a differential Bianchi identity following from the
$\xi^{M}$ gauge invariance of (\ref{actionprin}). First, we
need to
rewrite the gauge transformations.
Using (\ref{covdiv}) the transformation of the dilaton reads
 \be\label{dilvar}
  \delta_{\xi}e^{-2d} \ = \ \partial_{M}\big(e^{-2d}\xi^{M}\big) \ = \ e^{-2d}\nabla_{M}\xi^{M}\;.
 \ee
For the projector $P$ we have  $\delta_\xi P_{MN} = \widehat {\cal L}_\xi
P_{MN}$  because ${\cal H}$ transforms
with a generalized Lie derivative (\ref{gendiff}) and $\widehat{\cal L}_{\xi}\eta=0$ \cite{Hohm:2010pp}.  Due to the torsion constraint (\ref{GenTorsion}),  all partial derivatives in Lie derivatives
can be replaced by covariant derivatives.  We thus have
 \be\label{Pvar}
  \delta_{\xi}P_{MN} \ = \ \xi^{K}\nabla_{K}P_{MN}+2\nabla_{[ M}\xi_{K]}\,P^{K}{}_{N}+2\nabla_{[N}\xi_{K]}\,P^{K}{}_{M}\;.
 \ee
Using the covariant constancy of $P$ this becomes
 \be\label{Pvaroops}
  \delta_{\xi}P_{MN} \ = \ 2\, \nabla_{[ M}\xi_{\,\nin{N}]}\,\ +\ 2\, \nabla_{[N}\xi_{\,\nin{M}]}\,\;.
 \ee
Writing out the antisymmetrizations and using (\ref{sumtwo})
we have
\be
 \begin{split}
  \delta_{\xi}P_{MN} \ = \ & ~~\nabla_{M}\xi_{\nin{N}}-\nabla_{\nin{N}}\xi_{M}
  +   \nabla_{N}\xi_{\nin{M}}-\nabla_{\nin{M}}\xi_{N}\\[0.3ex]
   = \ & ~~ \nabla_{\nin{M}}\xi_{\nin{N}}+\nabla_{\bar{M}}\xi_{\nin{N}}
  -   \nabla_{\nin{N}}\xi_{\nin{M}} -\nabla_{\nin{N}}\xi_{\bar{M}}  \\
 & + \nabla_{\nin{N}}\xi_{\nin{M}}+\nabla_{\bar{N}}\xi_{\nin{M}}
  -   \nabla_{\nin{M}}\xi_{\nin{N}}-\nabla_{\nin{M}}\xi_{\bar{N}}  \\[0.3ex]
 = \   & ~~\nabla_{\bar{M}}\xi_{\nin{N}} -\nabla_{\nin{N}}\xi_{\bar{M}}
+\nabla_{\bar{N}}\xi_{\nin{M}}-\nabla_{\nin{M}}\xi_{\bar{N}} \;.
 \end{split}
 \ee
We can now write separate gauge transformations with respect to
${\xi}_{\,\nin{M}}$ and $\xi_{\bar{M}}$
 \be\label{covPvar}
 \begin{split}
  \delta_{\underline{\xi}}P_{MN} \ &= \  ~ \nabla_{\bar{M}}\,\xi_{\,\nin{N}}+\nabla_{\bar{N}}\,\xi_{\,\nin{M}} \;, \\
  \delta_{\bar{\xi}}P_{MN} \ &= \ -\big(\nabla_{\nin{M}}\xi_{\bar{N}}+\nabla_{\nin{N}}\xi_{\bar{M}}\big)\;.
 \end{split}
 \ee
For the dilaton we have, from (\ref{dilvar}),
 \be\label{dilvar99}
 \begin{split}
  \delta_{\underline{\xi}}\,e^{-2d} \ =& \ e^{-2d}\,\nabla_{\nin{M}}\xi^{\,\nin{M}}\,, \\
 \delta_{\bar \xi}\,e^{-2d} \ =& \ e^{-2d}\, \nabla_{\bar M}\xi^{\bar M}\,.
  \end{split}
 \ee

 Consider now the gauge variation  $\delta_{\underline{\xi}}$ of the
 action  (\ref{actionprin}).
Recalling that the curvature itself does not need to be varied
because it contributes only total derivatives, as in (\ref{GenrVarrr}),
we have
 \be\label{bianchiVar}
 \begin{split}
  0 \ &= \ \delta_{\underline{\xi}} \int  dx d\tilde{x}\,e^{-2d}P^{MK}P^{NL}
  \widehat {\cal R}_{MNKL} \\
  \ &= \ \int  dx d\tilde{x}\,e^{-2d}\left(\nabla_{\nin{P}}\xi^{\nin{P}}\,{\cal R}
  +2\big(\nabla^{\bar{M}}\xi^{\nin{K}}+\nabla^{\bar{K}}\xi^{\nin{M}}
  \big)P^{NL}\widehat {\cal R}_{MNKL}\right)\\
  \ &= \ -\int  dx d\tilde{x}\,e^{-2d} \xi^{\,\nin{P}}\left(\nabla_{\underline{P}}{\cal R}+4\nabla^{\bar{M}} {\cal R}_{\bar{M}\,\nin{N}\,\nin{P}}{}^{\,\nin{N}}\right)\;,
 \end{split}
 \ee
where we also used the property (\ref{EqualRic}).
The last contraction is (minus) the Ricci tensor, and since (\ref{bianchiVar}) holds for arbitrary $\xi^{\, \nin{P}}$ we conclude
 \be
  \nabla_{\nin{P}}{\cal R}\,-\, 4\,\nabla^{\bar{M}}{\cal R}_{\,\nin{P}\bar{M}} \ = \  0\;.
 \ee
Using the invariance under $\delta_{\bar{\xi}}$
 we get a similar looking equation with an opposite relative sign:
 \be
  \nabla_{\bar{P}}{\cal R}\, +\, 4\, \nabla^{\nin{M}} {\cal R}_{\nin{M}\bar{P}} \ = \ 0\;.
 \ee
These are the differential Bianchi identities of double field theory \cite{Siegel:1993th,Kwak:2010ew,Hohm:2010xe}.
We have not been able to find an {\em uncontracted}
differential Bianchi identity for
the full Riemann tensor, and we suspect that such an identity does not exist.
In fact, it is not hard
 to convince oneself that
 the naive Bianchi identity $\nabla_{[M}{\cal R}_{NK]PQ}=0$ does not hold
 by writing it out
in terms of connections.
As a further check, it is also straightforward
 to see that the double contraction of this naive Bianchi identity would give rise to an invalid contracted differential Bianchi identity.

\section{Riemann-squared  and the generalized metric}

Here we will investigate if
there exist manifestly $O(D,D)$
invariant terms quartic in derivatives and written with the generalized metric that,
for $b=\phi=0$, reduce to
the square of the Riemann tensor in some T-duality frame.
First, we work out the square of the Riemann tensor
in terms of the metric $g$.
 Then we identify one tensor structure that
cannot be reproduced from a generalized metric expression,
answering the above question in the negative.

\subsection{Outline of the approach}

Our results of the previous section indicate that natural steps do not
yield a  physical Riemann tensor in double field theory. They give
a four-index generalized tensor ${\cal R}_{MNPQ}$ that is not fully
determined in terms of the physical fields, but whose contractions
give physical  scalar and Ricci curvatures that were expected to exist.
It seems unlikely that there is a way to define a physical  ${\cal R}_{MNPQ}$ that is an $O(D,D)$ tensor, a generalized tensor, and reduces
to the Riemann tensor for particular combinations of indices.

We will  show in this section
that the
Riemann-squared scalar, familiar in $\alpha'$
corrections to the low-energy effective action of string theory,
cannot be obtained from a T-duality covariant expression built with
the generalized metric and the dilaton. More explicitly, we claim that
there is no $O(D,D)$ scalar ${\cal I}({\cal H}, d)$ such that it reduces
to Riemann squared when we set $\tilde \partial= 0$,  set the antisymmetric field $b_{ij}$
to zero, and set the dilaton $d$ to the value that corresponds to $\phi=0$
in the relation $e^{-2d} = \sqrt{-g} e^{-2\phi}$, namely $d = d_* \equiv -{1\over 2} \ln \sqrt{-g} $.   In other words, the answer to the following
question is negative:
\be
\label{the_big_question}
\hbox{Is there an}~O(D,D) ~\hbox{scalar} ~~{\cal I} ({\cal H}, d) ~~\hbox{such that}~~\,R_{ijkl} R^{ijkl} \ = \  {\cal I}  ({\cal H}, d)\Bigl|_{\tilde \partial = 0,\,b_{ij}=0, \,d= d_*} ~?
\ee
This happens because certain tensor
structures  appearing in the square of the Riemann tensor cannot
be reproduced from $O(D,D)$ invariant terms. This is a strong result, for the obstruction occurs just
by demanding that ${\cal I}$ be an $O(D,D)$ scalar.  An
${\cal I}({\cal H}, d)$ useful for double field theory
would also have to be a generalized scalar.
This result implies that even if there was a physical ${\cal R}_{MNPQ}$
that is both an $O(D,D)$ and a generalized tensor, and contained
components that give the Riemann tensor (after setting
$\tilde \partial = 0,\,b_{ij}=0, \,d= d_*$) it could not be of use in
constructing Riemann squared:
$O(D,D)$ contractions
would lead to canceling contributions.

Using the curvature scalar ${\cal R}$ and the Ricci tensor
${\cal R}_{\,\nin{M}\bar{N}}$,  both of which are
$O(D,D)$ tensors  and generalized  tensors,
 we can extend the double field theory action
 by the addition of higher-derivative terms with arbitrary
 coefficients $a$ and $b$:
 \be\label{DFTcorrection}
  S_{\rm DFT}\; \longrightarrow \;S_{\rm DFT}+\alpha^{\prime}\int dxd\tilde{x}\,e^{-2d}\big(a{\cal R}^2+b{\cal R}^{\,\nin{M}\bar{N}}{\cal R}_{\,\nin{M}\bar{N}}\big)\;.
 \ee
Setting $\tilde{\partial}=0$ this reduces to terms containing the square of the
conventional Ricci tensor and Ricci scalar. It is known, however,  that in string theory
also higher powers of the full Riemann tensor enter,
and therefore (\ref{DFTcorrection}) is not general enough
to first order in $\alpha^{\prime}$.   A correction $\Delta S$ proportional to Riemann
squared in the low energy action would take the form 
\be
\Delta S \ \sim \ \alpha' \int  dx \sqrt{-g}  e^{-2\phi} R_{ijkl} R^{ijkl}
 \ = \ \alpha' \int  dx\,  e^{-2d} R_{ijkl} R^{ijkl}\,.
\ee
If there had been an ${\cal I} ({\cal H}, d)$ that satisfies (\ref{the_big_question}) the term
\be
\Delta S_{\rm DFT} \ \sim \ \alpha' \int  dxd\tilde x  \, e^{-2d} ~{\cal I} ({\cal H},d)\,,
\ee
would have provided a suitable double field theory
extension (if ${\cal I}$ was also
a generalized tensor).   In the absence of ${\cal I} ({\cal H}, d)$
we can entertain some other possibilities.  It may be that a variant of
(\ref{the_big_question}) holds up to terms that can be dropped from
an action because they are total derivatives:
\be
e^{-2d} R_{ijkl} R^{ijkl} \ = \  {\cal I}  ({\cal H}, d)\Bigl|_{\tilde \partial = 0,\,b_{ij}=0} \ + \ \partial_i (\cdots ) \,.
\ee
We will not explore
this possibility in here, but it seems unlikely to work.  It seems to us
more likely that $\alpha'$ corrections require modifying the definition
of the generalized metric, as will be explained in the discussion
section.

\subsection{Terms quadratic in the Riemann tensor}
In this section we will compute the terms appearing
in the square of the Riemann tensor that are relevant for
the comparison with the generalized metric formulation
to be discussed in the next subsection.
In our conventions, which follow the book by Dirac \cite{Dirac}, the
Riemann tensor with all indices lowered is given by
 \be\label{RiemannLow}
R_{ijkl} \ = \ {1\over 2} \bigl(  \partial_j \partial_k g_{il} -\partial_i\partial_k g_{jl}- \partial_j \partial_l g_{ik} + \partial_i \partial_l g_{jk} \bigr)
+ \Gamma_{pil} \Gamma^p{}_{jk}  - \Gamma_{pik} \Gamma^p{}_{jl} \;,
\ee
with the Christoffel symbols
\be
\Gamma_{ijk} \ = \ {1\over 2} ( \partial_k g_{ij} + \partial_j g_{ik} - \partial_i g_{jk})\;, \qquad  \Gamma^i_{\, jk} \ \equiv \ g^{ip} \Gamma_{pjk} \,.
\ee
We also write (\ref{RiemannLow}) as
 \be\label{Riem0+}
  R_{ijkl} \ = \   R^0_{ijkl}+2\Gamma_{pi[l}\Gamma^{p}{}_{k]j}\;,
 \ee
where
  \be\label{R0}
  R^0_{ijkl} \ \equiv \ 2\partial_{[j}\partial_{[\underline{k}}\,g_{i]\underline{l}]}\;,
 \ee
with $[ab] \equiv {1\over 2} (ab - ba)$, and we have underlined indices
in order to avoid ambiguities in antisymmetrizations.
$R^0$ shares the symmetries of the full Riemann tensor,
 \be\label{R0symm}
  R^0_{ijkl} \ = \ -R^0_{jikl} \ = \ -R^0_{ijlk} \ = \ R^0_{klij}\;.
 \ee

Let us now consider the square of the Riemann tensor,
 \be\label{RiemSq}
(\hbox{Riem})^2 \ \equiv \   R_{ijkl} R^{ijkl}
 \ = \ R_{ijkl} \, g^{ir} g^{js}  g^{kt}  g^{lu}  \, R_{rstu}\;.
\ee
From the definition (\ref{RiemannLow}) we infer that this square
contains three different structures that are schematically
  \be\label{skbvgn}
(\partial \partial g_{**} )^2  \;, \qquad (\partial \partial g_{**}) (\partial g_{**})^2 \;, \qquad  (\partial g_{**})^4  \;.
 \ee
In order to establish our result it is sufficient to work out the
first two. We will see that while the first structures can be reproduced
from $O(D,D)$ invariant terms,
this is not so for the second structures, proving that the full square of the
Riemann tensor cannot be reproduced from an $O(D,D)$ invariant expression.

We begin by computing the terms of the first type, $(\partial \partial g_{**} )^2$,
from (\ref{RiemSq}).
Since terms involving the square of Christoffel symbols always
contain the structure $(\partial g_{**} )^2$, the terms we consider here
originate only from the square of $R^0$.
With (\ref{R0}) we then have\\[0.5ex]
  \be
 \begin{split}
 (\hbox{Riem})^2\Bigl|_{(\partial \partial g_{**} )^2}
 \; \ = \ \;  &2  \partial_{[j} \partial_{[\underline{k}}
g_{i]\,\underline{l}]} g^{ir} g^{js}  g^{kt}  g^{lu} \, R^0_{rstu}
\ = \ 2 \, \partial_{j} \partial_{k}
g_{il}~ g^{ir} g^{js}  g^{kt}  g^{lu} \, R^0_{rstu}\\
\ = \ \; &\partial_{j} \partial_{k}
g_{il}~ g^{ir} g^{js}  g^{kt}  g^{lu} \bigl( \partial_s \partial_t g_{ru}
- \partial_r \partial_t g_{su}  - \partial_s \partial_u g_{rt}
+ \partial_r \partial_u g_{st} \bigr)   \\[1.0ex]
\ = \ \;&\partial_{j} \partial_{k}
g_{il}~ g^{ir} g^{js}  g^{kt}  g^{lu} ~\partial_s \partial_t g_{ru}
~-~ 2 \, \partial_{j} \partial_{k}
g_{il}~ g^{ir} g^{js}  g^{kt}  g^{lu} ~ \partial_r \partial_t g_{su}
\\[1.0ex]
& +  \partial_{j} \partial_{k}
g_{il}~ g^{ir} g^{js}  g^{kt}  g^{lu} ~\partial_r \partial_u g_{st}\;,
\end{split}
 \ee\\[0.5ex]
where we combined in the third line two terms using the
symmetry properties of $g$ and of second partial derivatives.
After relabeling indices, this reads
  \be
  \label{kcbvg}
 \begin{split}
(\hbox{Riem})^2\Bigl|_{(\partial \partial g_{**} )^2}
 \; \ = \ \;  &g^{ij}   g^{kl}  g^{mn}g^{pq} ~ \partial_{i} \partial_{k}
g_{mp}~\partial_j \partial_l g_{nq}
-2\,  g^{ij} g^{kl}  g^{mn}  g^{pq} ~\partial_{k} \partial_{m}
g_{ip}~ \partial_j \partial_n g_{lq}
\\
&+  g^{ij}   g^{kl}  g^{mn}g^{pq} ~ \partial_{i} \partial_{k}
g_{mp}~ \partial_n \partial_q g_{jl} \\
\; \ = \ \;
&
g^{ij}   g^{kl}  g^{mn}g^{pq} \Bigl(
  \partial_{i} \partial_{k}
g_{mp}\,\partial_j \partial_l g_{nq}
 -2\,\partial_{k} \partial_{m}
g_{ip}\,\partial_j \partial_n g_{lq}
+\partial_{i} \partial_{k}
g_{mp}\, \partial_n \partial_q g_{jl}\Bigr)
\;.
\end{split}
 \ee

Let us now turn to the second structure in (\ref{skbvgn}).
It originates from cross terms of $R^{0}$ and the $\Gamma^2$ term
in (\ref{Riem0+}). Thus,
 \be
 \begin{split}
   (\hbox{Riem})^2\Bigl|_{(\partial \partial g_{**}) (\partial g_{**})^2}  \ &= \
   4R^{0\,nikm}\Gamma_{rnm}\Gamma^{r}{}_{ik} \\
   \ &= \
  R^{0\,nikm}\left(\partial_{m}g_{rn}+\partial_{n}g_{rm}-\partial_{r}g_{nm}\right)g^{rs}
  \left(\partial_{i}g_{ks}+\partial_{k}g_{is}-\partial_{s}g_{ik}\right)\;.
 \end{split}
 \ee
Using the symmetries (\ref{R0symm}) we can exchange $m\leftrightarrow n$ and
$i\leftrightarrow k$ simultaneously. It then follows that several terms combine, giving
\be\label{RSq112}
 \begin{split}
   (\hbox{Riem})^2\Bigl|_{(\partial \partial g_{**}) (\partial g_{**})^2}  \ = \
   & R^{0\,nikm}g^{rs}\Big(2\,\partial_m g_{rn}\,\partial_i g_{ks}+2\,\partial_m g_{rn}\,\partial_k g_{is}\\
   &-2\,\partial_m g_{rn}\,\partial_s g_{ik}-2\,\partial_r g_{nm}\,\partial_i g_{ks}
   +\partial_r g_{nm}\,\partial_s g_{ik}\Big)\;.
 \end{split}
 \ee

\subsection{Obstructions on the generalized metric formulation}

We attempt now to write $O(D,D)$ invariant expressions in terms
of the generalized metric that reproduce the above structures
(\ref{kcbvg}) and (\ref {RSq112}) when
setting $\tilde{\partial}^i=0$ and $b_{ij} = 0$.
In this situation
the generalized metric reads
\be\label{specialH}
{\cal H}_{MN} \ = \ \begin{pmatrix} {\cal H}^{ij} & {\cal H}^i_{~j} \\[0.5ex]
{\cal H}_i^{~j} & {\cal H}_{ij} \end{pmatrix} \ = \
\begin{pmatrix} g^{ij} &0 \\
0 & g_{ij} \end{pmatrix}\;.
\ee
Specifically, we will see that the only candidate $O(D,D)$ invariant expression
that could reproduce a certain tensor structure in the square of
the Riemann tensor is actually zero as a consequence of the
group properties of ${\cal H}_{MN}$.

We start with the $(\partial \partial g_{**} )^2$ terms in (\ref{kcbvg}). It turns out that they are
reproduced by  a  term ${\cal I}^{(2,2)}({\cal H})$
defined by
\be\label{22Hansatz}
\begin{split}
 {\cal I}^{(2,2)}({\cal H}) \ \equiv \ &-{1\over 2} \, {\cal H}^{IJ} {\cal H}^{KL} \, \partial_I \partial_K {\cal H}^{PQ}
\,\partial_J\partial_L   {\cal H}_{PQ} \\
 &+ 2\,  {\cal H}^{IJ} {\cal H}^{KL} \, \partial_I \partial_K {\cal H}^{MN}\,
\partial_J\partial_M {\cal H}_{LN}
+  \partial_I \partial_J {\cal H}^{KL}~
\partial_K\partial_L   {\cal H}^{IJ}\;.
\end{split}
\ee
The superscripts on ${\cal I}$ indicate the derivative structure of the
terms:  they are the product of a factor with two derivatives and another
factor with two derivatives.
In order to evaluate the reduction of
${\cal I}^{(2,2)}({\cal H})$ we set $\tilde{\partial}^i=0$ and insert (\ref{specialH}).
First note that any derivative must have a lower
index, $\partial_K \to  \partial_k$, and the index contracted with this derivative also becomes~$k$:
\be\label{22Hansatz99}
\begin{split}
 {\cal I}^{(2,2)} \ =  \ &-{1\over 2} \, {\cal H}^{ij} {\cal H}^{kl} \, \partial_i \partial_k {\cal H}^{PQ}
\,\partial_j\partial_l   {\cal H}_{PQ} \\
 &+ 2\,  {\cal H}^{ij} {\cal H}^{kL} \, \partial_i \partial_k {\cal H}^{mN}\,
\partial_j\partial_m {\cal H}_{LN}
+  \partial_i \partial_j {\cal H}^{kl}~
\partial_k\partial_l   {\cal H}^{ij}\;.
\end{split}
\ee
Because of the diagonal form of ${\cal H}$ in (\ref{specialH})
any mixed-index structure ${\cal H}^{iK}$ will only receive contributions
when $K$ is an upper lowercase index, giving ${\cal H}^{ik}$, with the
$k$ also appearing elsewhere as a lower index.
  Thus in the second term
above we can simply replace $L\to l$ and $N\to n$.  In the first term
there are two contributions for $P$ and $Q$, one with
structure ${\cal H}^{pq} \cdots {\cal H}_{pq}$ and the other
 ${\cal H}_{pq} \cdots {\cal H}^{pq}$.  Both turn out to give the same
 answer and we therefore have:
  \be
\label{sckvmvg}
{\cal I}^{(2,2)}(g) ~=~  - g^{ij} g^{kl} \, \partial_i \partial_k g^{pq} \, \partial_j
\partial_l  g_{pq}
~+~ 2 g^{ij} g^{kl} \, \partial_i \partial_k g^{mn} \,
\partial_j \partial_m g_{l n}
   ~+ ~\partial_i \partial_j g^{kl} \, \partial_k \partial_l g^{ij}\,.~
\ee
We can now  transform  the double derivatives
of upper-indexed metrics to derivatives of lower-indexed metrics using
\be
\label{sgbvgn}
\begin{split}
\partial_i \partial_j g^{-1}  &~=~ -\partial_i (g^{-1} \partial_j g g^{-1}) \\
&~=~g^{-1} (\partial_i g) g^{-1}(\partial_j g) g^{-1}  - g^{-1} (\partial_i
\partial_j g)  g^{-1}  + g^{-1} (\partial_j g) g^{-1}(\partial_i g) g^{-1}\,.
\end{split}
\ee
In components this reads
 \be
  \partial_{k}\partial_{l}g^{ij} \ = \ -g^{ir}g^{js}\partial_k \partial_l g_{rs}
  +2g^{ip}g^{jq}g^{rs}\partial_{(\underline{k}}g_{pr}\,\partial_{\underline{l})}g_{qs}\;.
 \ee
We use this in (\ref{sckvmvg}) and collect only the terms with two derivatives on
$g$:
\be
\begin{split}
{\cal I}^{(2,2)}(g)\Bigl|_{(\partial\partial g_{**})^2}  \, \ = \ \,  &g^{ij} g^{kl} \,( g^{ps} \partial_i \partial_k g_{st}g^{tq} )\, \partial_j
\partial_l  g_{pq}  -2 g^{ij} g^{kl} \, ( g^{ms}\partial_i \partial_k g_{st} g^{tn})\,
\partial_j \partial_m g_{l n}  \\
&+ (g^{ks}\partial_i \partial_j g_{st} g^{tl}) \, (g^{iu}\partial_k \partial_l g_{uv}g^{vj} )\,.~
\end{split}
\ee
After a straightforward relabeling of indices
one can compare with (\ref{kcbvg}) and confirm that
\be
~~{\cal I}^{(2,2)}(g)\Bigl|_{(\partial\partial g_{**})^2}
\ = \ (\hbox{Riemann})^2\Bigl|_{(\partial \partial g_{**} )^2} \;.
\ee
This shows that the proposed generalized metric combination (\ref{22Hansatz})
correctly reproduces the portion of (Riemann)$^2$ with two derivatives
on each field. But, as we will see, it does not produce all
of (Riemann)$^2$.

Let us now consider the $(\partial \partial g_{**}) (\partial g_{**})^2$ terms.
Note that  (\ref{sgbvgn}) implies  that ${\cal I}^{(2,2)}$ produces already several
terms of this type.  It is convenient to begin again with (\ref{sckvmvg})
to do this systematically.
Converting one of the $\partial^2 g^{**}$
metrics in the last term of ${\cal I}^{(2,2)}$ in (\ref{sckvmvg}) into $g_{**}$, we find
 \be
  \begin{split}
   {\cal I}^{(2,2)}(g) \ = \ & -g^{ik}\,g^{kl}\,\partial_{i}\partial_{k}g^{pq}\big(\partial_{j}\partial_{l}g_{pq}
   -2\partial_{j}\partial_{p}g_{lq}+\partial_{p}\partial_{q}g_{jl}\big)\\
   &+2g^{ij}g^{kl}g^{rs}\,\partial_{i}\partial_{k}g^{pq}\,\partial_{p}g_{jr}\partial_{q}g_{ls}\;.
  \end{split}
 \ee
The terms in parenthesis
 are proportional to $R^0$ and thus we conclude
  \be
  {\cal I}^{(2,2)}(g) \ = \ -2\,\partial_{i}\partial_{k}g^{pq}\,g^{ij}g^{kl}R^0_{qjlp}
  +2
 \,g^{ij}g^{kl}g^{rs} \,\partial_{i}\partial_{k}g^{pq}
   \partial_{p}g_{jr}\partial_{q}g_{ls}\;.
 \ee
The second term in here is produced by
minus $ {\cal I}^{(2,1,1)} ({\cal H}) $,
defined by
 \be\label{I211}
  {\cal I}^{(2,1,1)}({\cal H}) \ \equiv \ -2{\cal H}^{IJ}{\cal H}^{KL}{\cal H}^{RS}\partial_{I}\partial_{K}{\cal H}^{PQ}
  \partial_{P}{\cal H}_{JR}\,\partial_{Q}{\cal H}_{LS}\;.
 \ee
We can therefore write
 \be \label{kbb}
   {\cal I}^{(2,2)}(g)+ {\cal I}^{(2,1,1)}(g) \ = \ -2\partial_{i}\partial_{k}g^{pq}\,g^{ij}g^{kl}R^0_{qjlp}\;.
 \ee
We have shown that the terms on the right-hand side of this equation
are reproduced from the generalized metric expression ${\cal I}^{(2,2)}+ {\cal I}^{(2,1,1)}$.

We next investigate how much
the right-hand side of (\ref{kbb})
differs from the
square of the Riemann tensor.
For this purpose we first convert the
leftover
 $\partial^2g^{**}$ in (\ref{kbb})
into $g_{**}$,
 \be
  {\cal I}^{(2,2)}+{\cal I}^{(2,1,1)} \ = \  2g^{pm}g^{qn}\partial_{i}\partial_{k}g_{mn}\,g^{ij}g^{kl}R^0_{qjlp}
  -4g^{pm}g^{qn}g^{rs}\partial_{(\underline{i}}g_{mr}\partial_{\underline{k})}g_{ns}\,g^{ij}g^{kl}R^0_{qjlp}\;.
 \ee
The $\partial^2g_{**}$ structure in the first term inherits the antisymmetries from $R^0$ and so
this term is actually $(R^0)^2$.  Thus,
 \be\label{step782}
  \begin{split}
    {\cal I}^{(2,2)}+{\cal I}^{(2,1,1)} \ = \
    g^{ip}g^{jq}g^{km}g^{ln}R_{ijkl}^0R_{pqmn}^0
    -2R^{0\,nikm}g^{rs}\left(\partial_{i}g_{mr}\,\partial_{k}g_{ns}+\partial_{k}g_{mr}\,\partial_{i}g_{ns}\right)\;.
  \end{split}
 \ee
In here, the first term gives precisely the $(\partial \partial g_{**} )^2$ terms, as discussed above,
while the second
one gives some of the $(\partial \partial g_{**}) (\partial g_{**})^2$ terms. Comparing these with
the actual terms of this type appearing in the square of the Riemann tensor (\ref{RSq112}) finally implies
  \be\label{sstep9911}
  \begin{split}
    R^{ijkl}R_{ijkl} \ &= ~~{\cal I}^{(2,2)}(g)+{\cal I}^{(2,1,1)}(g)
    \\[1.0ex]
    &-R^{0\,nikm}g^{rs}\big(-2\partial_{k}g_{mr}\,
    \partial_{i}g_{ns}-2\partial_{m}g_{rn}\,\partial_{k}g_{is}-\partial_{r}g_{mn}\,\partial_{s}g_{ik}
    +4\partial_{m}g_{rn}\,\partial_{s}g_{ik} \big) \\[1.0ex]
    &+ {\cal O} ((\partial g)^4)\,.
  \end{split}
 \ee
The first line is reproduced by the generalized metric expressions (\ref{22Hansatz}) and (\ref{I211}).
We will now carefully examine the terms in the second line.
We will  identify one structure that cannot be written in terms
of the generalized metric.

We first expand
 \be
  \begin{split}
    -R&^{0\,nikm}g^{rs}
    \big(
   -2\partial_{k}g_{mr}\,
    \partial_{i}g_{ns}-2\partial_{m}g_{rn}\,\partial_{k}g_{is}-\partial_{r}g_{mn}\,\partial_{s}g_{ik}
    +4\partial_{m}g_{rn}\,\partial_{s}g_{ik} \big) \\[1.0ex]
   & = \ -\frac{1}{2}g^{np}g^{iq}g^{kl}g^{mt}g^{rs}
   \left(\partial_q\partial_l g_{pt} -\partial_p\partial_l g_{qt}
   -\partial_q\partial_t g_{pl}+\partial_p\partial_t g_{ql}\right) \\[0.5ex]
   &\qquad\qquad\qquad\qquad\quad  \left(-2\partial_{k}g_{mr}\,
    \partial_{i}g_{ns}-2\partial_{m}g_{rn}\,\partial_{k}g_{is}-\partial_{r}g_{mn}\,\partial_{s}g_{ik}
    +4\partial_{m}g_{rn}\,\partial_{s}g_{ik}\right) \\[1.0ex]
   & = \  -\frac{1}{2}g^{np}g^{iq}g^{kl}g^{mt}g^{rs}
   \big(
   -2\partial_{k}g_{mr}\,\partial_{i}g_{ns}\,\partial_q\partial_l g_{pt}
   +2\partial_{k}g_{mr}\,\partial_{i}g_{ns}\,\partial_p\partial_l g_{qt}
   \\&\qquad\qquad\qquad\qquad\qquad\;
   +2\partial_{k}g_{mr}\,\partial_{i}g_{ns}\,\partial_q\partial_t g_{pl}
    -2\partial_{k}g_{mr}\,\partial_{i}g_{ns}\,\partial_p\partial_t g_{ql}
    \\[1.0ex]
   &\qquad\qquad\qquad\qquad\qquad\;
   -2\partial_{m}g_{rn}\,\partial_{k}g_{is}\,\partial_q\partial_l g_{pt}
    +2\partial_{m}g_{rn}\,\partial_{k}g_{is}\,\partial_p\partial_l g_{qt}
     \\
   &\qquad\qquad\qquad\qquad\qquad\;
   +2\partial_{m}g_{rn}\,\partial_{k}g_{is}\,\partial_q\partial_t g_{pl}
    -2\partial_{m}g_{rn}\,\partial_{k}g_{is}\,\partial_p\partial_t g_{ql}
    \\[1.0ex]
   &\qquad\qquad\qquad\qquad\qquad\;
   -\partial_{r}g_{mn}\,\partial_{s}g_{ik} \,\partial_q\partial_l g_{pt}
    +\partial_{r}g_{mn}\,\partial_{s}g_{ik} \,\partial_p\partial_l g_{qt}
    \\
   &\qquad\qquad\qquad\qquad\qquad\;
   +\partial_{r}g_{mn}\,\partial_{s}g_{ik} \,\partial_q\partial_t g_{pl}
    -\partial_{r}g_{mn}\,\partial_{s}g_{ik} \,\partial_p\partial_t g_{ql}
   \\[1.0ex]
   &\qquad\qquad\qquad\qquad\qquad\;
   +4\partial_{m}g_{rn}\,\partial_{s}g_{ik} \,\partial_q\partial_l g_{pt}
    -4\partial_{m}g_{rn}\,\partial_{s}g_{ik} \,\partial_p\partial_l g_{qt}
    \\
   &\qquad\qquad\qquad\qquad\qquad\;
   -4\partial_{m}g_{rn}\,\partial_{s}g_{ik} \,\partial_q\partial_t g_{pl}
    +4\partial_{m}g_{rn}\,\partial_{s}g_{ik} \,\partial_p\partial_t g_{ql}  \;\big)\;.
  \end{split}
 \ee
Several terms in here can be combined,
 \be\label{finalexp}
  \begin{split}
   & = \  -\frac{1}{2}g^{np}g^{iq}g^{kl}g^{mt}g^{rs}
   \big(
   -2\uuline{\partial_{k}g_{mr}\,\partial_{i}g_{ns}\,
   \partial_q\partial_l g_{pt}}
   +8\,\partial_{k}g_{mr}\,\partial_{i}g_{ns}\,\partial_p\partial_l g_{qt}
    \\&\qquad\qquad\qquad\qquad\qquad\;
    -2\partial_{k}g_{mr}\,\partial_{i}g_{ns}\,\partial_p\partial_t g_{ql}
    \\[1.5ex]
   &\qquad\qquad\qquad\qquad\qquad\;
   -2\,\underline{\partial_{r}g_{mn}\,\partial_{s}g_{ik} \,
   \partial_q\partial_l g_{pt}}
    +2\,\partial_{r}g_{mn}\,\partial_{s}g_{ik} \,\partial_p\partial_l g_{qt}
    \\[1.5ex]
   &\qquad\qquad\qquad\qquad\qquad\;
   +4\partial_{m}g_{rn}\,\partial_{s}g_{ik} \,\partial_q\partial_l g_{pt}
    -4\partial_{m}g_{rn}\,\partial_{s}g_{ik} \,\partial_p\partial_l g_{qt}
    \\&\qquad\qquad\qquad\qquad\qquad\;
   -4\partial_{m}g_{rn}\,\partial_{s}g_{ik} \,\partial_q\partial_t g_{pl}
    +4\underline{\partial_{m}g_{rn}\,\partial_{s}g_{ik} \,\partial_p\partial_t g_{ql}}
     \\
   &\qquad\qquad\qquad\qquad\qquad\;
   -4\partial_{m}g_{rn}\,\partial_{k}g_{is}\,\partial_q\partial_l g_{pt}
    \;\big)\;,
  \end{split}
 \ee
where we have grouped the terms according to the index structure of the first
$\partial_{*}g_{**}$ factor.

In (\ref{finalexp}) we have underlined three terms that deserve special consideration.
All other terms can be reproduced by simply
replacing metrics for generalized metrics and partial derivatives by
$O(D,D)$ covariant
partial derivatives.  This happens because
the indices on derivatives (that are lower,
lowercase, when $\tilde \partial =0$) are contracted in such a way that
they force all other indices to become lowercase once we recall that
the generalized metric is diagonal.
To make this point more transparent, consider the second term
in (\ref{finalexp}):
\be
\label{2term}
-4\,g^{np}g^{iq}g^{kl}g^{mt}g^{rs}
\,\,\partial_{k}g_{mr}\,\partial_{i}g_{ns}\,\partial_p\partial_l g_{qt}\,.
\ee
Its $O(D,D)$ covariant extension is simply
 \be\label{EXAMple}
  -4\, {\cal H}^{NP} {\cal H}^{IQ} {\cal H}^{KL} {\cal H}^{MT} {\cal H}^{RS}\,\,\partial_{K}
  {\cal H}_{MR}\,\partial_{I}{\cal H}_{N S}\,\partial_{P}\partial_{L}{\cal H}_{QT}\;.
 \ee
 To see that this works we just follow the indices on
 derivatives (which must be lower, lowercase)
 and how they
 force indices to become lowercase.  From $\partial_K$ we
 have $K \to L$,  that is, we get $k, l$.  From the second derivative
 we get $I \to Q \to T \to M \to R\to S\to N \to P$, and all indices
 become as in (\ref{2term}).    One can readily check that the
 same happens for all other non-underlined terms.

Let us now consider the underlined terms in (\ref{finalexp}).
We start with the two terms with a single underline, which
take the form
\be
X_1(g) \ = \ g^{np}g^{iq}g^{kl}g^{mt}g^{rs} \,\partial_{r}g_{mn}\,\partial_{s}g_{ik} \,
   \partial_q\partial_l g_{pt}
    \ - \ 2 g^{np}g^{iq}g^{kl}g^{mt}g^{rs}\,\partial_{m}g_{rn}\,\partial_{s}g_{ik} \,\,\partial_p\partial_t g_{ql}  \,.
\ee
On each of the terms, each of the $\partial g$ factors can be
transformed into derivatives of inverse metrics via the identity
$\partial g^{-1} = - g^{-1} (\partial g)  g^{-1}$,
 \be\label{probstruc}
X_1(g) \ = \    g^{rs}\,\partial_r g^{pt}\,\partial_s g^{ql}\,\partial_q\partial_l g_{pt}
 \ -\ 2\, g^{mt}\, \partial_m g^{ps}\, \partial_s g^{ql}
 \,\, \partial_p\partial_t g_{ql}\;.
 \ee
 These two structures can also be reproduced from an expression
 in terms of ${\cal H}$, with due care to double counting and extra
 terms that are to be thought
 of as higher order.  We claim that the following is the answer
 \be
{\cal I}_{X_1}({\cal H}) \ = \    \frac{1}{2}{\cal H}^{RS}\partial_{R}{\cal H}^{PT}\,\partial_{S}{\cal H}^{QL}\,\partial_{Q}\partial_{L}{\cal H}_{PT} \
    -\  {\cal H}^{MT}\partial_M {\cal H}^{PS}\partial_S {\cal H}^{QL}\partial_P\partial_T {\cal H}_{QL}\,.
  \ee
The first step in the reduction gives
 \be
{\cal I}_{X_1}\ = \    \frac{1}{2}\,g^{rs}\partial_{r}{\cal H}^{PT}\,
\partial_{s}g^{ql}\,\partial_{q}\partial_{l}{\cal H}_{PT} \
    -\  g^{mt}\partial_m g^{ps}\partial_s {\cal H}^{QL}\partial_p\partial_t {\cal H}_{QL}\,.
  \ee
This time we are left with contractions that give rise to two terms each,
 \be
 \begin{split}
{\cal I}_{X_1}(g)\ = \  & \phantom{-} \ \frac{1}{2}\,g^{rs}\,\partial_{r}g^{ql}\,
\partial_{s}g^{pt}\partial_{q}\partial_{l}g_{pt} +
 \ \frac{1}{2}\,g^{rs}\,\partial_{r}g^{ql}\,
 \partial_{s}g_{pt}\,\partial_{q}\partial_{l}g^{pt}  \\   \
  &  -\  g^{mt}\partial_m g^{ps}\,\partial_s g^{ql}\partial_p\partial_t g_{ql} \   -\  g^{mt}\partial_m g^{ps}\, \partial_s g_{ql}\,\partial_p\partial_t g^{ql}\,.
 \end{split}
  \ee
Using (\ref{sgbvgn})
one can readily see that the second term on each line
equals the first,  up to  $(\partial g)^4$ terms. Thus, we have
\be
{\cal I}_{X_1}(g) \ = \ X_1 (g) + {\cal O} ((\partial g)^4)\,.
\ee
This shows that the terms with a single underline can be reproduced
using the generalized metric, up to $(\partial g)^4$ terms
that must be considered once all $\partial^2 g (\partial g)^2$ terms are under control.

Let us finally consider the double-underlined term in (\ref{finalexp}), which turns out
to be problematic.  The term is
\be\label{problmterm}
Z(g) \ = \  g^{np}g^{iq}g^{kl}g^{mt}g^{rs}
  \,\,  \partial_{k}g_{mr}\,\partial_{i}g_{ns}\,
   \partial_q\partial_l g_{pt}\ = \
    -g^{np}g^{iq}g^{kl}\partial_k g^{ts}\ \partial_i g_{ns}\partial_q\partial_l g_{pt}\;,
\ee
where we rewrote the leftmost $\partial g$ in terms of $\partial g^{-1}$.
The only candidate $O(D,D)$ invariant term that could reproduce this structure is proportional to
 \be
 {\cal I}_Z \ = \  {\cal H}^{NP}  {\cal H}^{IQ}{\cal H}^{KL}\partial_{K}{\cal H}^{TS}\partial_{I}{\cal H}_{NS}\,
  \partial_{Q}\partial_{L}{\cal H}_{PT}\;.
 \ee
The claim is that $ {\cal I}_Z$ is in fact
 zero up to terms $(\partial {\cal H})^4$ --- which in turn give rise to structures
involving $(\partial g)^4$
and are thus of different type. To see this we raise and lower indices using $\eta$ on the
one hand and
the analogue of (\ref{sgbvgn}) for ${\cal H}$
on the other:
 \be
  \begin{split}
{\cal I}_Z \ &= \   \underline{{\cal H}^{NP}} \,  {\cal H}^{IQ} {\cal H}^{KL}\partial_{K} \underline{{\cal H}^{TS}}\, \partial_{I}\underline{{\cal H}_{NS}}\,
  \partial_{Q}\partial_{L}\underline{{\cal H}_{PT}} \\[1.0ex]
  \ &= \
  {\cal H}_{NP}   {\cal H}^{IQ}{\cal H}^{KL}\partial_{K}{\cal H}_{TS}
  \partial_{I}{\cal H}^{NS}
  \underline{\partial_{Q}\partial_{L}{\cal H}^{PT}} \\[1.0ex]
  \ &= \ - \underline{{\cal H}_{NP}}  \,
  {\cal H}^{IQ}{\cal H}^{KL}
  \underline{{\cal H}^{PR}}\,
  {\cal H}^{TM}
  \partial_{K}{\cal H}_{TS}\,
  \underline{\partial_{I}{\cal H}^{NS}}\,
  \partial_{Q}\partial_{L}{\cal H}_{RM}\,+\,(\partial {\cal H})^4\\[1.0ex]
  \ &= \   -{\cal H}^{TM}{\cal H}^{KL}{\cal H}^{IQ}\partial_{I}{\cal H}^{RS}\partial_{K}{\cal H}_{TS}\,
  \partial_{Q}\partial_{L}{\cal H}_{RM} \,+\,(\partial {\cal H})^4\\[1.0ex]
   \ &= \   -{\cal H}^{NP}{\cal H}^{IQ}{\cal H}^{KL}\partial_{K}{\cal H}^{TS}\partial_{I}{\cal H}_{NS}
  \partial_{L}\partial_{Q}{\cal H}_{TP} \,+\,(\partial {\cal H})^4\ \\[1.0ex]
  \ &=  \ - {\cal I}_Z
  + \,(\partial {\cal H})^4\;.
 \end{split}
 \ee
As help to the reader, the underlined factors in each term
denote those factors that
participate in the simplification leading to the next term.
In the step before the last line we relabeled indices ($ I \leftrightarrow K, \,
Q \leftrightarrow L,\,  R \to T \to N,  \, M \to P$).
 Thus, up to  $(\partial {\cal H})^4$ terms, this structure
is minus itself and thus zero.

One may wonder if
the dilaton $d$ can be used to help
reproduce the above problematic structure.
Unfortunately, this is not the case. Rather, the role of the dilaton can be understood as follows.
Whenever a tensor contains the
 structure $g^{kl}\partial_{m}g_{kl}$, the generalized metric cannot be used to
reproduce it. This follows because the corresponding $O(D,D)$ invariant term is minus itself by its group
properties and thus vanishes:
 \be
  {\cal H}^{KL}\partial_{M}{\cal H}_{KL} \ = \ -{\cal H}_{KL}\partial_{M}{\cal H}^{KL} \ = \ - {\cal H}^{KL}\partial_{M}{\cal H}_{KL}
  \ \equiv \ 0\;.
 \ee
In the first step we recalled that ${\cal H}^{KL}$ is the inverse of ${\cal H}_{KL}$, and in the second
step we raised and lowered indices with the constant $\eta_{MN}$.
In order to reproduce the structure $g^{kl}\partial_{m}g_{kl}$
we can use the $O(D,D)$ invariant dilaton $d$.
Since
$e^{-2d}=\sqrt{g}e^{-2\phi}$ we have, for $\tilde{\partial}^{i}=0$,
 \be\label{dilder}
  \partial_{M}d
   \;\;\, \rightarrow \;\;\, \partial_{m}d \ = \ \partial_{m}\phi-\frac{1}{4}g^{kl}\partial_{m}g_{kl}\;.
 \ee
This means that
\be
(-4 \partial_M d)\Bigl|_{\tilde\partial = 0 ,
\phi = 0}  ~\to ~ g^{kl}\partial_{m}g_{kl}\, ,
\ee
provides the desired $O(D,D)$ covariantization of the structure.
In fact, the $O(D,D)$ invariant scalar curvature given in \cite{Hohm:2010pp} can be
systematically constructed as follows. Start with the scalar curvature of Riemannian geometry
written in terms of $g_{ij}$. For each term that can be reproduced
using
the generalized metric
include the corresponding $O(D,D)$ covariant term. All terms that cannot be reproduced from a
generalized metric expression turn out to contain the structure
$g^{kl}\partial_{m}g_{kl}$, which is covariantized by $(-4\partial_M d)$.
It can be checked that this covariantization of the Ricci scalar
gives the generalized scalar ${\cal R}$ constructed in \cite{Hohm:2010pp}
and discussed in this paper.
On the other hand, for the
problematic structure (\ref{problmterm})
the dilaton does not help, as it
contains no contractions of the $g^{kl}\partial_{m}g_{kl}$ type.
As a side remark we point out that since the dilaton is of no use in constructing the
T-duality invariant extension of the Riemann tensor-squared, this suggests that
in a field basis in which the first $\alpha^{\prime}$ correction consists only of the square of
the Riemann tensor, the dilaton itself does not receive higher-derivative corrections.
Intriguingly, this is confirmed by explicit computations in string theory
\cite{Print-86-0243(PRINCETON)}.

Let us point out that for low-dimensional toy models like $D=2$ there may exist additional
manipulations to rewrite the structure (\ref{problmterm}) such that
it can be reproduced from a generalized metric or dilaton expression. In fact,
in $D=2$ the Riemann tensor is fully
determined by the scalar curvature and so the square of the generalized scalar ${\cal R}$
must contain Riemann-square. Incidentally, note that according to our formula
for the number of undetermined connections given after (\ref{Fconnection}) all
connections are determined in $D=2$. In contrast,
it is clear that for general $D$
there are no additional identities that would allow for such manipulations.

Summarizing,
for general $D$
there is no $O(D,D)$ invariant expression
in terms of the generalized metric that reproduces the required structure
appearing in the square of the Riemann tensor.
As a result
there is no $O(D,D)$ invariant term fourth-order in derivatives that reproduces the square of the full Riemann tensor.

\section{Discussion: T-duality and $\alpha^{\prime}$ corrections}
In this paper we have investigated
the possible existence of
a double field theory Riemann tensor ${\cal R}_{MNPQ}$
satisfying conditions 1) -- 4) and (A),
as stated in the introduction.
In the first part of this paper  we gave
 a self-contained presentation of a  metric-like
 formalism
 introducing connections and invariant curvatures
along the lines of the  frame-like  approach of Siegel~\cite{Siegel:1993th}.
The main difference with the
related metric-like formalism
of Jeon, Lee, and Park~\cite{Jeon:2010rw}
is that we keep track of undetermined pieces in
the connection and their effects on curvatures.
Our analysis sheds new  light on the Riemann tensor.
Specifically, we showed that the
components that are fully determined in terms of the physical fields
vanish identically as a consequence of an algebraic Bianchi identity.
Thus, within this formalism,
there is no Riemann tensor meeting all conditions 1) -- 4).
There is a Riemann tensor satisfying conditions 1) -- 3).
It is
an $O(D,D)$ tensor,  a generalized tensor, and it determines
${\cal R}_{MN}$ and ${\cal R}$. It is not, however,  fully determined
in terms of the physical fields.
The components of ${\cal R}_{MNPQ}$
that are independent of
undetermined connections vanish.

In the second part of this paper
we investigated a related question.
We asked  if there is a four-derivative
 $O(D,D)$  invariant function
of the generalized metric
and the dilaton
that  reduces  in some T-duality frame (and with $b_{ij} = \phi =0$)
 to the square of the Riemann tensor.
We find that the answer is negative:
for general $D$ there is no $O(D,D)$ covariantization of
Riemann-square
in terms of the generalized metric and the dilaton.
Such covariantization, if it existed, could be used as a Lagrangian for higher-derivative terms in double field theory.
This result implies that even if a double field theory Riemann tensor satisfying conditions 1) -- 4) exists,
it could not provide a T-duality covariantization of Riemann-squared --
its square would have to be zero.

Let us now briefly discuss the significance of this result.
Suppose we had succeeded
in constructing an $O(D,D)$ invariant in terms of ${\cal H}_{MN}$
and $d$  that reduces to
the square of the Riemann tensor in some T-duality frame.
Then we would be able to write a  general action with
four derivatives as some arbitrary linear
combination of the squares of generalized
Riemann, generalized Ricci, and generalized scalar curvature.
{\em Any} of these actions would be exactly invariant under the original forms
of the T-duality and generalized diffeomorphisms that leave the
original two-derivative action invariant.
This would be unexpected,
for the field redefinitions
 \be
  g_{ij} \;\rightarrow\; g_{ij}+\alpha^{\prime}\left(a_1\, R_{ij}+a_2\, g_{ij} R \right)\, ,
 \ee
that respect diffeomorphism invariance,
map $\alpha'$-corrected actions into each other
in that they alter
the coefficients of Ricci-squared and $R$-squared terms.
After such  field redefinitions the T-duality transformation
of $g_{ij}$
will acquire $\alpha'$ corrections, in
conflict with the above implication of the
(hypothetical) existence of a physical
generalized Riemann tensor.

Useful insights into the structure of T-duality in double field theory
to order $\alpha'$
are suggested by
the
computations
of Meissner~\cite{Meissner:1996sa}.\footnote{Later
work of Kaloper and Meissner~\cite{Kaloper:1997ux} did not
use the generalized metric.  It evaluated  $\alpha'$ corrections
to T-duality transformations arising in backgrounds with one abelian isometry.}
He considered `cosmological' models, i.e., the reduction of gravitational actions with
higher-order corrections to one dimension.
The resulting theory can be written in an $O(D,D)$ invariant way only if the
formula for the generalized metric in terms of the $g$ and $b$ fields
receives $\alpha'$ corrections.  For double field theory such
a possibility would imply that the theory can be written in terms of
a generalized metric
$\overline{\cal H}_{MN} (g, b)$ of the form
 \be
 \overline{\cal H}_{MN}(g, b)  \ = \ {\cal H}_{MN}(g, b)
 \ +\ \alpha^{\prime}\,
  \delta{\cal H}_{MN}(g, b)  + {\cal O} (\alpha'^2) \;,
 \ee
where ${\cal H}(g, b)$ is the
generalized metric (\ref{firstH})
and  $\overline {\cal H}_{MN} (g, b)$ is a symmetric
$O(D,D)$ matrix  to order $\alpha'$.  Since (\ref{firstH}) is a general
parameterization of a symmetric $O(D,D)$ matrix, this means that one can
write
\be
 \overline {\cal H}_{MN} (g, b) \ = \ {\cal H}_{MN} (g', b')\,,
 \ee
  where
$(g',b')$ are $\alpha'$ corrected versions of $(g,b)$.
The results of~\cite{Meissner:1996sa}
(see eqs.~(4.11)--(4.12))
suggest a redefinition of the type 
 \be\label{Hredf}
 (g')^{ij}
 \ = \  g^{ij}  +   \alpha^{\prime}g^{ik}g^{jl}g^{pq}g^{rs}\big(a_1\, \partial_r g_{kp}\,\partial_{s}g_{lq}
  +a_2\,  \partial_r b_{kp}\,\partial_{s}b_{lq}\big)\;.
 \ee
It would be interesting to see if the problematic structure that we identified
in the square of the Riemann tensor can be removed
with such a  field redefinition.
Once the action is written in terms of $ {\cal H}_{MN} (g', b')$, one
could view $(g',b')$ as the new field variables with standard (uncorrected)
T-duality transformations.  The redefinition (\ref{Hredf}) does not preserve
manifest general covariance because it involves first derivatives of the metric rather than tensors.
Thus
generalized diffeomorphisms
would
receive $\alpha'$ corrections.
It would be interesting to see if the field basis suggested by string field theory
has to play a special role here (see \cite{Hohm:2011dz} for the explicit map between different
field variables).

While the  generalized Riemann tensor discussed in this paper
is not fully determined by the physical fields, we expect it to play
a crucial role in the construction of general T-duality invariant
$\alpha'$ corrections.
As discussed in section~\ref{thecompon} this tensor has components of type $(4,0), (3,1), (1,3),$ and
$(0,4)$:
 \be\label{freaklist-final}
  {\cal R}_{\nin{M}\nin{N}\nin{K}\nin{L}}\;,\quad
  {\cal R}_{\nin{M}\nin{N}\nin{K}\bar{L}}\;, \quad
 {\cal R}_{\nin{M}\bar{N}\bar{K}\bar{L}}\;, \quad
 {\cal R}_{\bar{M}\bar{N}\bar{K}\bar{L}}\;,
 \ee
 all of which depend on undetermined connections.  We believe that a suitable linear combination of squares of these curvatures will have
the property that the undetermined part can be removed by a field
redefinition.

It is amusing  to speculate on the meaning of our results for the
geometry that underlies string theory.  The absence of a physical
Riemann tensor seems to follow from the requirement of duality
covariance.  Since the Riemann tensor is needed for the construction
of the interactions in the theory, we are forced to learn how to work
with a partially physical, generalized Riemann tensor.  This is all we
seem to have.  In Riemannian geometry a spacetime is flat if and
only if the Riemannian curvature vanishes. In the absence of a
physical  Riemann tensor in string theory there would seem to be
no obvious way to characterize flat space!

\section*{Acknowledgments}
We would like to thank Ashoke Sen for collaboration at an
initial stage of this project and Igor Klebanov, Seung Ki Kwak, Silviu Pufu, and Dan Waldram for
helpful discussions.

This work is supported by the U.S. Department of Energy (DoE) under the cooperative
research agreement DE-FG02-05ER41360, the
DFG Transregional Collaborative Research Centre TRR 33
and the DFG cluster of excellence "Origin and Structure of the Universe".

 \appendix

\section{Relation to frame formalism}
\setcounter{equation}{0}
Here we explain the equivalence of the `metric-like' formalism discussed in this paper 
and the `frame-like' formalism of Siegel \cite{Siegel:1993th}, extending the discussion given 
in sec.~5.3 of \cite{Hohm:2010xe}. The  
vielbein $e_{A}{}^{M}$, with inverse $e_{M}{}^{A}$, 
carries an $O(D,D)$ index $M$ and a flat index $A$ 
 with respect to the local tangent space group $GL(D)\times GL(D)$.
This flat index splits as $A=(a,\bar{a})$, where $a$ refers to the left $GL(D)$ and $\bar{a}$ to the right $GL(D)$. 
In order to describe only physical fields the vielbein  $e_{A}{}^{M}$ needs to satisfy constraints that 
are written in terms of the tangent space metric 
${\cal G}$ defined by
 \be\label{tangentmetric}
  {\cal G}_{AB} \ \equiv \ e_{A}{}^{M}\,e_{B}{}^{N}\,\eta_{MN}\,,  ~~~
  \hbox{with inverse} ~~~  {\cal G}^{AB} \ = \  \eta^{MN} \,e_M{}^A\, e_N{}^B \;.
 \ee
Flat indices are raised and lowered with~${\cal G}$ while $O(D,D)$ indices
are raised and lowered with~$\eta$.   Moreover,  
$e_M{}^A  = \eta_{MN} {\cal G}^{AB} e_B{}^N$. 
We impose the constraints  
 \be\label{GConstr}
  {\cal G}_{a\bar{b}} \ = \ 0\;, \qquad   {\rm sig} ({\cal G}_{ab}) \ = \ (+\,-\,\ldots -)\;, \quad {\rm sig}({\cal G}_{\bar{a}\bar{b}}) \ = \ (-\,+\,\ldots +)\;, 
 \ee
where `${\rm sig}$'  
denotes  the signature. 
Note that the signatures of ${\cal G}_{ab}$  and 
${\cal G}_{\bar{a}\bar{b}}$  are opposite
in order to be consistent with the $(D,D)$ signature of ${\cal G}_{AB}$.
The assignment of signatures here 
complies with the conventions of  \cite{Hohm:2010xe}. By Sylvester's theorem of inertia, 
the constraints (\ref{GConstr}) are  
$GL(D)\times GL(D)$ invariant. 

The projectors $P$ and $\bar{P}$ and the generalized metric 
${\cal H}$   can be defined in 
terms of the frame field as in \cite{Hohm:2010xe}:
 \be
  P_{M}{}^{N} \ \equiv \ e_{aM}\,e^{aN}\;, \qquad \bar{P}_{M}{}^{N} \ \equiv \ e_{\bar{a}M}\,e^{\bar{a}N}\;, \qquad
  {\cal H}_{M}{}^{N} \ \equiv \ \frac{1}{2}\big(\bar{P}_{M}{}^{N}-P_{M}{}^{N}\big)\;. 
 \ee
As required, these projectors satisfy $P^2=P$, $\bar{P}^2=\bar{P}$ and, 
using the first constraint of (\ref{GConstr}), $P\bar{P}=0$.

Following Siegel we may now introduce spin connections $\omega_{ABC}$ for the local $GL(D)\times GL(D)$ symmetry and impose 
covariant constraints in order to determine (part of) them in terms of the physical fields. 
These spin connections then uniquely determine Christoffel connections by means of a vielbein postulate 
as follows. We introduce a covariant derivative $D$ with respect to the spin and Christoffel connection 
and postulate that the frame field $e_{A}{}^{M}$ is 
covariantly constant: 
 \be\label{vielbeinpost}
  D_{M}e_{A}{}^{N} \ \equiv \ \partial_{M}e_{A}{}^{N}+\Gamma_{MK}{}^{N}e_{A}{}^{K}+\omega_{MA}{}^{B}e_{B}{}^{N} \ = \ 0\;.
 \ee
Here 
$$\omega_{MA}{}^B  \ = \ e_M{}^C \omega_{CA}{}^B\,. $$ 
Note that  
\be
\label{ddelta}
D_M \delta_A{}^B \ = \ 0\,.
\ee
  Because of the factorized gauge group, the non-vanishing 
spin  
connections are $\omega_{Ma}{}^{b}$ and 
$\omega_{M\bar{a}}{}^{\bar{b}}$.   
The covariant derivative $D_{M}$ reduces to the covariant derivative $\nabla_{M}$ discussed in 
this paper when acting on tensors with only curved indices. 
Moreover, the covariant derivative 
 \be
 D_A \ \equiv \ e_A{}^M D_M 
\ee 
reduces to the flat covariant derivative $\nabla_A$  
of Siegel when acting on tensors with only $GL(D)\times GL(D)$ indices. Thus, with the 
vielbein being covariantly constant, any statement about `tangent space' objects can be translated into a statement about `world' objects 
and viceversa, in precise analogy to conventional Riemannian geometry. For instance, by (\ref{vielbeinpost}) the 
Christoffel connection is determined by the frame field and the spin connection according to 
 \be\label{GammaSol}
  \Gamma_{MNK} \ = \ -e_{M}{}^{A}e_{N}{}^{B}e_{K}{}^{C}\omega_{ABC}-e_{N}{}^{A}\partial_{M}e_{AK}\;.
 \ee
 In the following we will show that the constraints of Siegel 
imply via (\ref{vielbeinpost}) our constraints (1)--(4) on $\Gamma$ and thus that the frame formalism of Siegel 
is equivalent to the metric-like formalism discussed in this paper. 

The frame formulation imposes the following constraints on the spin connection:
\begin{itemize}
  \item[(i)] The tangent space metric (\ref{tangentmetric}) is covariantly constant, 
   \be
  \nabla_A {\cal G}_{BC} \ = \ 0\;.
   \ee
Since ${\cal G}_{BC}$ has only flat indices, the above implies that
 \be  
  D_{A}{\cal G}_{BC} \ = \ 0 ~~\to  ~~ D_{M}{\cal G}_{BC}\ = \ 0 \,.
   \ee
 Because of (\ref{vielbeinpost}) and (\ref{ddelta}), we have
 that $e_{M}{}^{A}$ is also covariantly constant and thus
 we can write
 \be
D_{M}\big(  e_N{}^B e_K{}^C {\cal G}_{BC}\big)\ = \ 0  \quad\to \quad 
D_{M}  \eta_{NK} \ = \ 0\,, 
 \ee
 by use of (\ref{tangentmetric}). 
Since $\eta$ only has $O(D,D)$ indices, the last equation
above implies  $\nabla_{M}\eta_{NK}=0$, which is constraint (1). 
  Moreover, we now readily derive the covariant constancy of 
  $P$, $\bar{P}$ and therefore of ${\cal H}$, thus implying constraint (3).  
  For example,
  \be
  \nabla_M P_N{}^K  \ =  \   D_M P_N{}^K  
  \ = \  D_M  (e_{aN} e^{aK}) \;, 
  \ee
  where in the last step we noted that when $D_M$ acts on an object
  $R_A{}^A$ with a contracted flat index there is no contribution
  from the spin connection.  Given the diagonal form of the 
  spin connection components the same is true for the action of $D_M$ on an object of the form $R_a{}^a$
  or $R_{\bar a} {}^{\bar a}$.  Thus we are allowed to use the full
  covariant derivative $D_M$ in the last expression above.  Since $D_M$
  is a derivation and the vielbeins are covariantly constant we conclude
  that $\nabla_M P_N{}^K = 0 $. 
  
 \item[(ii)]  The second constraint requires that in the C-bracket 
  \be
   \big[\xi_1,\xi_2\big]_{\rm C}^{M} \ \equiv \ \xi_1^{N}\partial_{N}\xi_2^{M}-\frac{1}{2}\xi_{1N}\partial^{M}\xi_{2}^{N}-(1\leftrightarrow 2)\;, 
  \ee
we can flatten the indices 
 by introducing covariant derivatives as follows,
   \be\label{covCbracket99}  
    \big[\xi_1,\xi_2\big]_{\rm C}^{A} \ \equiv 
     \ e_{M}{}^{A} \big[\xi_1,\xi_2\big]_{\rm C}^{M}  \ = \ \xi_1^{B}\nabla_{B}\xi_2^{A}-\frac{1}{2}\xi_{1B}\nabla^{A}\xi_{2}^{B}-(1\leftrightarrow 2)\;. 
  \ee
Since the derivatives act on flat indices we can replace $\nabla$ by $D$
and the constraint becomes
  \be\label{covCbracket}
    \big[\xi_1,\xi_2\big]_{\rm C}^{A} 
      \ = \ \xi_1^{B}D_{B}\xi_2^{A}-\frac{1}{2}\xi_{1B}D^{A}\xi_{2}^{B}-(1\leftrightarrow 2)\;. 
  \ee
 This constraint implies the generalized torsion constraint (2) in the form (\ref{GenTorsion}). In order to see this we 
recall that 
eqs.~(3.29)--(3.30) in \cite{Hohm:2010pp} show that the generalized Lie derivative can be written 
 in terms of the C-bracket as 
  \be
    \widehat{\cal L}_{\xi}V^{M} \ = \ \big[\xi,V\big]_{\rm C}^{M} +\frac{1}{2}\partial^{M}\big(V^{N}\xi_{N}\big) 
    \ = \ \big[ \xi,V\big]_{\rm C}^{A}\,e_{A}{}^{M}+\frac{1}{2}\nabla^{M}\big(V^{N}\xi_{N}\big)\;,  
  \ee
 where we used that the partial derivative of the scalar $V^{N}\xi_{N}$ coincides with the covariant derivative.   
 Inserting (\ref{covCbracket}) we obtain 
 \be
   \widehat{\cal L}_{\xi}V^{M}  =  \Bigl(
   \xi^B D_B V^A - V^B D_B \xi^A  - {1\over 2} \xi_B D^A V^B
   + {1\over 2} V_B 
   D^A \xi^B \Bigr)\,e_{A}{}^{M}+\frac{1}{2}\nabla^{M}\big(V^{N}\xi_{N}\big)  
   \,.   \ee
Using the covariant constancy of the vielbein  
and converting all indices into curved indices 
we can replace  $D$'s by $\nabla$'s and obtain 
  \be
  \begin{split}
   \widehat{\cal L}_{\xi}V^{M} \ &= \ \xi^{N}\nabla_{N}V^{M}-V^{N}\nabla_{N}\xi^{M}-\frac{1}{2}\xi_{N}\nabla^{M}V^{N}
   +\frac{1}{2}V_{N}\nabla^{M}\xi^{N}+\frac{1}{2}\nabla^{M}\big(V^{N}\xi_{N}\big)  \\
   \ &= \  \xi^{N}\nabla_{N}V^{M}+\big(\nabla^{M}\xi^{N}-\nabla^{N}\xi^{M}\big)V_{N} \ \equiv \ \widehat{\cal L}_{\xi}^{\;\nabla} V^M\;. 
  \end{split}
  \ee
 We recovered (\ref{GenTorsion}) and thus constraint (2), as we wanted to show. 
 
 \item[(iii)] The third constraint requires 
   \be\label{framedilatonconstr99}  
     \int e^{-2d}\,V\nabla_{A}V^{A} \ = \ -\int e^{-2d}\,V^{A}\nabla_{A}
     V\;.  
    \ee
We can replace $\nabla$ by $D$:
    \be\label{framedilatonconstr}
     \int e^{-2d}\,VD_{A}V^{A} \ = \ -\int e^{-2d}\,V^{A}D_{A}
     V\;. 
    \ee
On the right-hand side we can immediately pass to $O(D,D)$ indices.
On the left-hand side this requires use of  (\ref{vielbeinpost}).
We thus find
    \be\label{framedilatonconstr987}
     \int e^{-2d}\,VD_MV^{M} \ = \ -\int e^{-2d}\,V^{M}D_{M}
     V\;. 
    \ee
Replacing $D$ by $\nabla$, as is allowed now,  we obtain
(\ref{dilatonconstr}), thus implying constraint (4). 
Alternatively, the constraint can also be verified  
  explicitly by inserting eq.~(2.37) 
  of~\cite{Hohm:2010xe}   
   into the trace of (\ref{GammaSol}), from which we 
  recover~(\ref{Gammatraced}).

\end{itemize}

In total, the constraints (i)--(iii) of the frame formalism imply, via (\ref{vielbeinpost}), the constraints (1)--(4) of the metric-like formalism, 
thereby establishing the equivalence of both formulations.


\begin{thebibliography}{99}
\bibitem{Hull:2009mi}
  C.~Hull, B.~Zwiebach,
  ``Double Field Theory,''
  JHEP {\bf 0909}, 099 (2009).
  [arXiv:0904.4664 [hep-th]],
  ``The Gauge algebra of double field theory and Courant brackets,''
  JHEP {\bf 0909}, 090 (2009).
  [arXiv:0908.1792 [hep-th]].

\bibitem{Hohm:2010jy}
  O.~Hohm, C.~Hull and B.~Zwiebach,
  ``Background independent action for double field theory,''
  JHEP {\bf 1007} (2010) 016
  [arXiv:1003.5027 [hep-th]].

\bibitem{Hohm:2010pp}
  O.~Hohm, C.~Hull and B.~Zwiebach,
  ``Generalized metric formulation of double field theory,''
  JHEP {\bf 1008} (2010) 008
  [arXiv:1006.4823 [hep-th]].

\bibitem{Siegel:1993th}
  W.~Siegel,
  ``Superspace duality in low-energy superstrings,''
  Phys.\ Rev.\  D {\bf 48}, 2826 (1993)
  [arXiv:hep-th/9305073],
  ``Two vierbein formalism for string inspired axionic gravity,''
  Phys.\ Rev.\  D {\bf 47}, 5453 (1993)
  [arXiv:hep-th/9302036].

  \bibitem{Tseytlin:1990nb}
A.~A.~Tseytlin,
``Duality Symmetric Formulation Of String World Sheet Dynamics,''
Phys.\ Lett.\ B {\bf 242}, 163 (1990);
``Duality Symmetric Closed String Theory And Interacting Chiral Scalars,''
Nucl.\ Phys.\ B {\bf 350}, 395 (1991).

\bibitem{Duff:1989tf}
  M.~J.~Duff,
  ``Duality Rotations In String Theory,''
  Nucl.\ Phys.\ B {\bf 335}, 610 (1990),
  M.~J.~Duff and J.~X.~Lu,
  ``Duality Rotations In Membrane Theory,''
  Nucl.\ Phys.\ B {\bf 347}, 394 (1990).

\bibitem{Hohm:2010xe}
  O.~Hohm, S.~K.~Kwak,
  ``Frame-like Geometry of Double Field Theory,''
  J.\ Phys.\ A {\bf A44}, 085404 (2011).
  [arXiv:1011.4101 [hep-th]],

\bibitem{Kwak:2010ew}
  S.~K.~Kwak,
  ``Invariances and Equations of Motion in Double Field Theory,''
  JHEP {\bf 1010} (2010) 047
  [arXiv:1008.2746 [hep-th]].
\bibitem{Hohm:2011gs}
  O.~Hohm,
  ``T-duality versus Gauge Symmetry,''
  arXiv:1101.3484 [hep-th], \\
   B.~Zwiebach,
  ``Double Field Theory, T-Duality, and Courant Brackets,''
  [arXiv:1109.1782 [hep-th]].

\bibitem{Hohm:2011dz}
  O.~Hohm,
  ``On factorizations in perturbative quantum gravity,''
  JHEP {\bf 1104}, 103 (2011).
  [arXiv:1103.0032 [hep-th]].

\bibitem{Hohm:2011ex}
  O.~Hohm, S.~K.~Kwak,
  ``Double Field Theory Formulation of Heterotic Strings,''
  JHEP {\bf 1106}, 096 (2011).
  [arXiv:1103.2136 [hep-th]].

\bibitem{Hohm:2011zr}
  O.~Hohm, S.~K.~Kwak, B.~Zwiebach,
  ``Unification of Type II Strings and T-duality,''
  Phys.\ Rev.\ Lett.\  {\bf 107}, 171603 (2011),
    [arXiv:1106.5452 [hep-th]],
  ``Double Field Theory of Type II Strings,''
  JHEP {\bf 1109}, 013 (2011),
    [arXiv:1107.0008 [hep-th]].

\bibitem{arXiv1108.4937}
  O.~Hohm and S.~K.~Kwak,
  ``Massive Type II in Double Field Theory,''
  JHEP\ {\bf 1111} (2011) 086
  [arXiv:1108.4937 [hep-th]].

\bibitem{arXiv1111.7293}
  O.~Hohm and S.~K.~Kwak,
  ``N=1 Supersymmetric Double Field Theory,''
  arXiv:1111.7293 [hep-th].

\bibitem{Hillmann:2009ci}
  C.~Hillmann,
  ``Generalized E(7(7)) coset dynamics and D=11 supergravity,''
  JHEP {\bf 0903}, 135 (2009).
  [arXiv:0901.1581 [hep-th]].

\bibitem{Berman:2010is}
  D.~S.~Berman, M.~J.~Perry,
  ``Generalized Geometry and M theory,''
  JHEP {\bf 1106}, 074 (2011).
  [arXiv:1008.1763 [hep-th]],
  D.~S.~Berman, H.~Godazgar, M.~J.~Perry,
  ``SO(5,5) duality in M-theory and generalized geometry,''
  Phys.\ Lett.\  {\bf B700}, 65-67 (2011).
  [arXiv:1103.5733 [hep-th]],
  D.~S.~Berman, E.~T.~Musaev, M.~J.~Perry,
  ``Boundary Terms in Generalized Geometry and doubled field theory,''
  [arXiv:1110.3097 [hep-th]],
  D.~S.~Berman, H.~Godazgar, M.~Godazgar, M.~J.~Perry,
  ``The Local symmetries of M-theory and their formulation in generalised geometry,''
  [arXiv:1110.3930 [hep-th]],
  D.~S.~Berman, H.~Godazgar, M.~J.~Perry, P.~West,
  ``Duality Invariant Actions and Generalised Geometry,''
  [arXiv:1111.0459 [hep-th]].

\bibitem{West:2010ev}
  P.~West,
  ``$E_{11}$, generalised space-time and IIA string theory,''
  Phys.\ Lett.\  {\bf B696}, 403-409 (2011).
  [arXiv:1009.2624 [hep-th]], \\
  A.~Rocen, P.~West,
  ``E11, generalised space-time and IIA string theory: the R-R sector,''
  [arXiv:1012.2744 [hep-th]].

\bibitem{Jeon:2010rw}
  I.~Jeon, K.~Lee, J.~-H.~Park,
  ``Differential geometry with a projection: Application to double field theory,''
  JHEP {\bf 1104}, 014 (2011).
  [arXiv:1011.1324 [hep-th]].

\bibitem{Jeon:2011cn}
  I.~Jeon, K.~Lee, J.~-H.~Park,
  ``Stringy differential geometry, beyond Riemann,''
  Phys.\ Rev.\  {\bf D84}, 044022 (2011).
  [arXiv:1105.6294 [hep-th]].

\bibitem{Jeon:2011vx}
  I.~Jeon, K.~Lee, J.~-H.~Park,
  ``Incorporation of fermions into double field theory,''
  JHEP {\bf 1111}, 025 (2011).
  [arXiv:1109.2035 [hep-th]],
  ``Supersymmetric Double Field Theory: Stringy Reformulation of Supergravity,''
  arXiv:1112.0069 [hep-th].

\bibitem{Schulz:2011ye}
  M.~B.~Schulz,
  ``T-folds, doubled geometry, and the SU(2) WZW model,''
  [arXiv:1106.6291 [hep-th]].

\bibitem{Copland:2011yh}
  N.~B.~Copland,
  ``Connecting T-duality invariant theories,''
  Nucl.\ Phys.\  {\bf B854}, 575-591 (2012).
  [arXiv:1106.1888 [hep-th]],
  ``A Double Sigma Model for Double Field Theory,''
  [arXiv:1111.1828 [hep-th]].

\bibitem{Thompson:2011uw}
  D.~C.~Thompson,
  ``Duality Invariance: From M-theory to Double Field Theory,''
  JHEP {\bf 1108}, 125 (2011).
  [arXiv:1106.4036 [hep-th]].

\bibitem{Albertsson:2011ux}
  C.~Albertsson, S.~-H.~Dai, P.~-W.~Kao, F.~-L.~Lin,
  ``Double Field Theory for Double D-branes,''
  JHEP {\bf 1109}, 025 (2011).
  [arXiv:1107.0876 [hep-th]].

\bibitem{Andriot:2011uh}
  D.~Andriot, M.~Larfors, D.~Lust, P.~Patalong,
  ``A ten-dimensional action for non-geometric fluxes,''
  JHEP {\bf 1109}, 134 (2011).
  [arXiv:1106.4015 [hep-th]],
  G.~Aldazabal, W.~Baron, D.~Marques, C.~Nunez,
  ``The effective action of Double Field Theory,''
  JHEP {\bf 1111}, 052 (2011).
  [arXiv:1109.0290 [hep-th]],
  D.~Geissbuhler,
  ``Double Field Theory and N=4 Gauged Supergravity,''
  [arXiv:1109.4280 [hep-th]].

\bibitem{Coimbra:2011nw}
  A.~Coimbra, C.~Strickland-Constable, D.~Waldram,
  ``Supergravity as Generalised Geometry I: Type II Theories,''
    [arXiv:1107.1733 [hep-th]],
  ``$E_{d(d)} \times \mathbb{R}^+$ Generalised Geometry, Connections and M theory,''
  arXiv:1112.3989 [hep-th].

\bibitem{arXiv:0710.2719}
  M.~Gualtieri,
  ``Branes on Poisson varieties,''
  arXiv:0710.2719 [math.DG].

\bibitem{Batalin:2007xi}
  I.~A.~Batalin and K.~Bering,
  ``Odd Scalar Curvature in Field-Antifield Formalism,''
  J.\ Math.\ Phys.\  {\bf 49}, 033515 (2008)
  [arXiv:0708.0400 [hep-th]].
  I.~A.~Batalin and K.~Bering,
  ``A Comparative Study of Laplacians and Schrodinger-Lichnerowicz-Weitzenbock Identities in Riemannian and Antisymplectic Geometry,''
  J.\ Math.\ Phys.\ \ {\bf 50}, 073504  (2009)
  [arXiv:0809.4269 [hep-th]].


\bibitem{Meissner:1996sa}
  K.~A.~Meissner,
  ``Symmetries of higher order string gravity actions,''
  Phys.\ Lett.\  {\bf B392}, 298-304 (1997).
  [hep-th/9610131].

\bibitem{Kaloper:1997ux}
  N.~Kaloper and K.~A.~Meissner,
  ``Duality beyond the first loop,''
  Phys.\ Rev.\ D {\bf 56}, 7940 (1997)
  [hep-th/9705193].

\bibitem{Dirac}
P.A.M.~Dirac, ``General Theory of Relativity,"
Princeton University Press.

\bibitem{Print-86-0243(PRINCETON)}
  C.~G.~Callan, Jr., I.~R.~Klebanov and M.~J.~Perry,
  ``String Theory Effective Actions,''
  Nucl.\ Phys.\ B\ {\bf 278}, 78  (1986).




\end{thebibliography}
\end{document}